\documentclass[traditabstract,onecolumn]{aa} 
\usepackage{amsmath,graphicx,natbib,eufrak,txfonts}

\newcommand{\mr}{\mathrm}

\newcommand{\be}{\begin{equation}}
\newcommand{\ee}{\end{equation}}
\newcommand{\bfmath}[1]{\mbox{\boldmath${\mr #1}$\unboldmath}}

\newcommand{\beq}{\begin{eqnarray}}
\newcommand{\eeq}{\end{eqnarray}}
\newcommand{\ensav}[1]{\left\langle #1 \right\rangle}

\begin{document}
\title{Weak lensing power spectra for precision cosmology}
\subtitle{Multiple-deflection, reduced shear and lensing bias corrections}
\author{Elisabeth Krause \and Christopher M. Hirata}
\authorrunning{E. Krause \& C. M. Hirata}
\institute{Caltech M/C 350-17, Pasadena, CA 91125, USA}
\date{}

\abstract{
It is usually assumed that the ellipticity power spectrum measured in weak lensing observations can be expressed as an integral over the underlying matter power 
spectrum. This is true at order ${\mathcal O}(\Phi^2)$ in the gravitational potential.  We extend the standard calculation, constructing all corrections to order 
${\mathcal O}(\Phi^4)$. There are four types of corrections: corrections to the lensing shear due to multiple-deflections; corrections due to the fact that shape 
distortions probe the reduced shear $\gamma/(1-\kappa)$ rather than the shear itself; corrections associated with the non-linear conversion of reduced shear to mean 
ellipticity; and corrections due to the fact that observational galaxy selection and shear measurement is based on galaxy brightnesses and sizes which have been 
(de)magnified by lensing.  We show how the previously considered corrections to the shear power spectrum correspond to terms in our analysis, and highlight new terms 
that were not previously identified.  All correction terms are given explicitly as integrals over the matter power spectrum, bispectrum, and trispectrum, and are 
numerically evaluated for the case of sources at $z=1$.  We find agreement with previous works for the ${\mathcal O}(\Phi^3)$ terms.  We find that for ambitious 
future surveys, the ${\mathcal O}(\Phi^4)$ terms affect the power spectrum at the $\sim 1-5\sigma$ level; they will thus need to be accounted for, but are 
unlikely to represent a serious difficulty for weak lensing as a cosmological probe.
}

\keywords{cosmology: theory -- gravitational lensing -- large scale structure of the Universe}

\maketitle

\section{Introduction}

Cosmic shear, the distortion of light from distant galaxies by the tidal gravitational field of the intervening large scale structure, is an excellent tool to probe 
the matter distribution in the universe. The statistics of the image distortions are related to the statistical properties of the large scale matter distribution and 
can thereby be used to constrain cosmology. Current results already demonstrate the power of cosmic shear observations at constraining the clustering amplitude 
$\sigma_8$ and the matter density $\Omega_{\mr m}$ \citep[e.g.,][]{Cosmos,Tim,CFHTLS, Fu}. Furthermore, cosmic shear provides an ideal tool to study dark energy 
through measuring the evolution of non-linear structure with large future surveys (DES\footnote{www.darkenergysurvey.org}, LSST\footnote{www.lsst.org},JDEM\footnote{http://jdem.gsfc.nasa.gov}, Euclid\footnote{http://sci.esa.int/science-e/www/object/index.cfm?fobjectid=42266}). These upcoming large weak lensing experiments will limit the 
statistical uncertainties to the percent level.

In order to extract cosmological information from these cosmic shear experiments, the increased data quality needs to be accompanied by a thorough treatment of 
systematic errors. On the observational side, this requires accurate information on the redshift distribution of source galaxies \citep{MHH_photoz} and precise measurements of galaxy 
shapes which correct for observational systematics such as pixelization, noise, blurring by seeing and a spatially variable point spread function \citep[see][]{STEP2,GREAT08}. On the theoretical 
side, astrophysical contaminants, like source lens clustering \citep{BvWM97,Schneider02}, intrinsic alignment \citep{King_ic} and the correlation between the gravitational shear and 
intrinsic ellipticities of galaxies \citep{HS04, King_il,JS08,Zhang08,JS09}, need to be understood and removed.
The prediction of lensing observables also requires precise models of the non-linear matter power spectrum and models for the relation between lensing distortion 
and large scale matter distribution which go beyond linear theory. While N-body simulations may predict the non-linear dark matter power spectrum with percent 
level accuracy in the near future \citep{Coyote1, Coyote2}, the effect of baryons, which is a significant contamination to the weak lensing signal above $l\sim 2000$ 
\citep{Jing06, Rudd08}, is more difficult to account for and is the subject of ongoing work.

In this paper, we consider corrections to the relation between the observed lensing power spectra and the 
non-linear matter density field.
In the regime of weak lensing, the observed galaxy ellipticities ($e_I$) are an estimator of the reduced shear $g_I = 
\gamma_I/(1-\kappa)$,
\be
\ensav{e_I}  = C\frac{\gamma_I}{1-\kappa}\; ,
\label{eq:e}
\ee
where $C$ is a constant which depends on the type of ellipticity estimator \citep[e.g.][]{SS95, SS97} and the properties of the galaxy population under consideration, $\gamma_I$ is a component of the shear, $\kappa$ is the convergence, and the subscript $I$ refers to the two components of the 
ellipticity/shear \citep[see e.g. ][for more details]{BS01}. The two-point statistics of the measured ellipticities are simply related to the reduced shear power 
spectrum. \citet{CH} have calculated the shear power spectrum to fourth order in the gravitational potential. For the reduced shear power spectrum there exists an 
approximation to third order in the gravitational potential \citep{DSW}. \citet{Shapiro08} has demonstrated that on angular scales relevant for dark energy parameter 
estimates the difference between shear and reduced shear power spectra is at the percent level and ignoring these corrections will noticeably bias dark energy 
parameters inferred from future weak lensing surveys.

\citet{FS_lens} introduced another type of corrections, termed \emph{lensing bias}, which has a comparable effect on the shear power spectrum as the reduced shear 
correction: Observationally, shear is only estimated from those galaxies which are bright enough and large enough to be identified and to measure their shape. This 
introduces cuts based on observed brightness and observed size, both of which are (de)magnified by lensing \citep[e.g.][]{BTP_mag, Jain_size}, and will thus bias the 
sampling of the cosmic shear field.

In the following we complete the calculation of the reduced shear power spectrum to fourth order in the gravitational potential to include multiple deflections and to 
account for the effects of lensing bias and the non-linear conversion between ellipticity and reduced shear.  We consider all lensing-related effects through 
${\mathcal O}(\Phi^4)$, but do {\em not} include effects associated with the sources (source clustering and intrinsic alignment corrections).

This paper is organized as follows: We describe our technique for calculating higher order lensing distortions and power spectra in Sect.~\ref{sec:method}. 
Derivations of the different types of corrections to the shear and reduced shear power spectra are given in Sect.~\ref{sec:shear} through Sect.~\ref{sec:lensbias}. We 
quantify the impact of these corrections on future surveys in Sect.~\ref{sec:impact} and discuss our results in Sect.~\ref{sec:discussion}.

\section{Calculational method}
In this section we derive the higher order lensing distortions following \citet{HS03}, and introduce our technique and notation for calculating power spectrum corrections.

Throughout this calculation we assume a flat universe and work in the flat sky approximation. We use a unit system based on setting the speed of light $c=1$, which makes potentials dimensionless. We use the Einstein summation convention and sum over all Roman indices appearing twice in a term. Lower case, italic type Roman indices $a,b,c,\, ... =1,2$ are used to for Cartesian components of two dimensional vectors and tensors; capital case, italic type Roman indices $I,J,K,\, ...=1,2$ are used for the components of polars which are defined with reference to a Cartesian coordinate system but have different transformation properties. Greek indices are used for redshift slices.

\label{sec:method}

\subsection{Lensing distortion tensor}
We work in the flat sky approximation and choose the sky to lie in the 
$xy$-plane. Photons travel roughly along the $- \hat{\mathbf z}$ direction and are deflected by the Newtonian potential $\Phi$ generated by the nonrelativistic matter inhomogeneities. As long as their deflection from the $- \hat{\mathbf z}$ direction is small, they observe a metric \citep[e.g.][]{HS03}
\be
ds^2 = a^2(\tau)\left[-(1+2\Phi)d\tau^2+(1-2\Phi)\left(d\chi^2 + \chi^2(d n_x^2+ d n_y ^2)\right)\right],
\ee
where $a$ is the scale factor, $\chi$ is the comoving radial distance, and $\mathbf n$ is the angular coordinate of the photon path on the sky.
We calculate the deflection angle of a light ray from its null geodesic equation
\be
\frac{\mr d}{\mr d \chi} \left( \frac{\mr d \mathbf n}{\mr d \chi} \chi \right) = -2\frac{\partial\Phi(\mathbf{x}(\mathbf{n},\chi);z(\chi))}{\partial\mathbf n} \chi\; ,
\label{eq:NGE}
\ee
where $\Phi(\mathbf{x}; z)$ is the Newtonian potential at position $\mathbf{x}$ and redshift $z$, with initial conditions $\mathbf n (\chi = 0) = \mathbf n_{0}$ and $\partial_{\chi} \mathbf n(\chi = 0) = 0$.

To first order in $\Phi$, the integration is performed along the unperturbed photon trajectory, this is the so-called Born approximation.
Taylor expanding Eq. (\ref{eq:NGE}) to third order in $\Phi$ we obtain a perturbative solution for the deflection angle $\mathbf d \equiv \mathbf n - \mathbf n_0$
\begin{align}
\nonumber n_i(z_{\mr s}) = & \;  {n_0}_i + {d}^{(1)}_i(z_{\mr s})+ {d}^{(2)}_i(z_{\mr s})+ {d}^{(3)}_i(z_{\mr s})\\ 
= & {n_0}_i + d^{(1)}(z_{\mr s})- 2\!\int_0^{\chi_{\mr s}}\!\!\!\mr d \chi W(\chi,\chi_{\mr s})\chi^2 \Phi_{,ia}(\chi) d_a^{(1)}(\chi) 
- 2\!\int_0^{\chi_{\mr s}}\!\!\!\mr d \chi W(\chi,\chi_{\mr s})\chi^2 \left(\frac{1}{2}\chi \Phi_{,iab}(\chi) d_a^{(1)}(\chi)d_b^{(1)}(\chi) +\Phi_{,ia}(\chi) d_a^{(2)}(\chi) \right)
\label{eq:Taylor}
\\
\nonumber  =& \;  {n_0}_i -2 \!\int_0^{\chi_{\mr s}}\!\!\! \mr d \chi W(\chi, \chi_{\mr s})\chi \Phi_{,i}(\chi) +4\!\int_0^{\chi_{\mr s}}\!\!\! \mr d \chi W(\chi, \chi_{\mr s})\chi^2\!\int_0^{\chi}\!\!\! \mr d \chi' W(\chi', \chi)\chi' \Phi_{,ia}(\chi)\Phi_{,a}(\chi') \\
\nonumber & - \left( 4\!\int_0^{\chi_{\mr s}}\!\!\! \mr d \chi W(\chi, \chi_{\mr s})\chi^3\!\int_0^{\chi}\!\!\! \mr d \chi' W(\chi', \chi)\chi' \!\int_0^{\chi}\!\!\! \mr d \chi'' W(\chi'', \chi)\chi'' \Phi_{,iab}(\chi)\Phi_{,a}(\chi')\Phi_{,b}(\chi'')\right.\\
&+\left. 8\!\int_0^{\chi_{\mr s}}\!\!\! \mr d \chi W(\chi, \chi_{\mr s})\chi^2\!\int_0^{\chi}\!\!\! \mr d \chi' W(\chi', \chi)\chi'^2\!\int_0^{\chi'}\!\!\! \mr d \chi'' W(\chi'', \chi')\chi''\Phi_{,ia}(\chi)\Phi_{,ab}(\chi')\Phi_{,b}(\chi'')\right),
\label{eq:n}
\end{align}
where $W(\chi',\chi) = \left(\frac{1}{\chi'}-\frac{1}{\chi}\right)\Theta(\chi - \chi')$ with $\Theta(x)$ the Heaviside step function. Here $\chi_{\mr s} = \chi(z_{\mr s})$ is the comoving distance of a source at 
redshift $z_{\mr s}$, commata represent comoving spatial transverse derivatives. These spatial derivatives are evaluated at the unperturbed position
$\Phi(\chi) = \Phi(\mathbf n_0 \chi,\chi;z(\chi))$ unless otherwise indicated.
The first and second order deflection angles are identical to those found by \citet{HS03}
\footnote{Our notation differs from \citet{HS03} in using spatial instead of angular derivatives to simplify comparison with \citet{CH, DSW, Shapiro08}}.
The third order deflection angles are caused by the two types of second order transverse displacement in the Taylor expansion of $\Phi(\mathbf x;z)$ shown in Eq. (\ref{eq:Taylor}). We discuss the difference between these terms after Eq. (\ref{eq:shear}).

The distortion of a light ray is then described by the Jacobian matrix
\be 
\mathbf A (\mathbf n_0, z_{\mr s}) =\frac{\partial \mathbf n (z_{\mr s})}{\partial \mathbf n_0} = \left(\begin{array}{cc}1-\kappa-\gamma_1 & -\gamma_2 -\omega\\-\gamma_2 +\omega&1-\kappa +\gamma_1\end{array}\right)\; ,
\label{eq:defa}
\ee
where $\gamma_I$ are the cartesian components of the shear, and $\omega$ induces an (unobservable) rotation of the image. Using (\ref{eq:n}), the distortion tensor $\psi_{ij} = \delta_{ij} - A_{ij}$ is given by
\be
\psi_{ij}(\mathbf{n}_0,z_{\mr s})  = \psi_{ij}^{(1)}(\mathbf{n}_0,z_{\mr s})+\psi_{ij}^{(2)}(\mathbf{n}_0,z_{\mr s})+\psi_{ij}^{(3A)}(\mathbf{n}_0,z_{\mr 
s})+\psi_{ij}^{(3B)}(\mathbf{n}_0,z_{\mr s})+\psi_{ij}^{(3C)}(\mathbf{n}_0,z_{\mr s}),
\label{eq:shear0}
\ee
where
\begin{align}
\nonumber 
\psi_{ij}^{(1)}(\mathbf{n}_0,z_{\mr s}) =&\;2\!\int_0^{\chi_{\mr s}}\!\!\! \mr d \chi W(\chi, \chi_{\mr s})\chi^2  \Phi_{,ij}(\chi)\;, \\
\nonumber
\psi_{ij}^{(2)}(\mathbf{n}_0,z_{\mr s}) =&-4\!\int_0^{\chi_{\mr s}}\!\!\! \mr d \chi W(\chi, \chi_{\mr s})\chi^2\!\int_0^{\chi}\!\!\! \mr d \chi' W(\chi', \chi)\chi'\left\{ 
\Phi_{,ia}(\chi)\chi'\Phi_{,aj}(\chi') + \chi \Phi_{,ija}(\chi)\Phi_{,a}(\chi')\right\}\;, \\
\nonumber  
\psi_{ij}^{(3A)}(\mathbf{n}_0,z_{\mr s}) =
 &+ 4\!\underbrace{\int_0^{\chi_{\mr s}}\!\!\! \mr d \chi W(\chi, \chi_{\mr s})\chi^4\!\int_0^{\chi}\!\!\! \mr d \chi' W(\chi', \chi)\chi' \!\int_0^{\chi}\!\!\! \mr d \chi'' W(\chi'', \chi)}_{\Longrightarrow \chi' < \chi,\; \chi''< \chi}
\chi'' \Phi_{,ijab}(\chi)\Phi_{,a}(\chi')\Phi_{,b}(\chi'')\;,\\
\nonumber  
\psi_{ij}^{(3B)}(\mathbf{n}_0,z_{\mr s}) =
 &+ 8\!\underbrace{\int_0^{\chi_{\mr s}}\!\!\! \mr d \chi W(\chi, \chi_{\mr s})\chi^3\!\int_0^{\chi}\!\!\! \mr d \chi' W(\chi', \chi)  \chi^{\prime 2} \!\int_0^{\chi}\!\!\! \mr d 
\chi'' W(\chi'', \chi)}_{\Longrightarrow \chi' < \chi,\; \chi''< \chi}
\chi'' \Phi_{,iab}(\chi) 
\Phi_{,aj}(\chi')\Phi_{,b}(\chi'')\;, {\rm~and}\\
\psi_{ij}^{(3C)}(\mathbf{n}_0,z_{\mr s}) =
 & +8\!\underbrace{
\int_0^{\chi_{\mr s}}\!\!\! \mr d \chi W(\chi, \chi_{\mr s})\chi^2 \!\int_0^{\chi}\!\!\! \mr d \chi' W(\chi', \chi)\chi^{\prime 2}\!\int_0^{\chi'}\!\!\! \mr d \chi'' W(\chi'', \chi')}_{\Longrightarrow \chi'' <\chi' < \chi}
\chi'' \frac{\partial}{\partial n_{0,j}}[\Phi_{,ia}(\chi)\Phi_{,ab}(\chi')\Phi_{,b}(\chi'')]\;,
\label{eq:shear}
\end{align}
where we have used the symmetry of the integrals over $\chi'$ and $\chi''$ in the derivation of $\psi_{ij}^{(3B)}$.
This calculation automatically includes the ``Born correction'' and ``lens-lens coupling'' corrections considered by \citet{CH}. Compared to their approach, we find 
additional terms $\psi_{ij}^{(3C)}$ which give the third order corrections caused by three lenses placed at different locations along the line of sight ($ \chi'' 
<\chi' < \chi$), namely the derivatives of the last term in Eq. (\ref{eq:shear}). These include the two terms previously considered by \citet{sc06}, however, we will show 
in Sect.~\ref{sec:shear} that within the Limber approximation, the 3C term does not contribute to the shear power spectrum at ${\mathcal O}(\Phi^4)$.

The convergence, shear, and rotation are expressible in terms of $\psi_{ij}$ by the usual rules $\kappa=\frac12(\psi_{11}+\psi_{22})$,
$\gamma_1=\frac12(\psi_{11}-\psi_{22})$, $\gamma_2=\frac12(\psi_{12}+\psi_{21})$, and $\omega=\frac12(\psi_{12}-\psi_{21})$.

Note that while our derivation of the deflection angle is based on the small angle approximation $d\ll 1$, in the flat sky approximation the elements of the distortion matrix need not be as small.

\subsection{Fourier space: first order}
Since we work in terms of power spectra, we need to transform these equations to Fourier space. In the flat-sky approximation,
\be
\psi_{ij}(\mathbf{n}_0,z_{\mr s}) = \int\frac{\mr d^2\mathbf l}{(2\pi)^2} \tilde\psi_{ij}(\mathbf l, z_{\mr s}) \mr e^{\mr i\mathbf l\cdot\mathbf n_0}.
\ee
The angular cross power spectra of two fields $\Gamma$ and $\Gamma'$ is then defined by
$\langle\tilde\Gamma(\mathbf l)\tilde\Gamma'(\mathbf l')\rangle = (2\pi)^2C_{\Gamma\Gamma'}(l)\delta_{\mr D}(\mathbf l+\mathbf l')$ with $\delta_{\mr D}$ the Dirac delta function, which has units $[\delta_{\mr D}(\mathbf x)] = [x]^{-n}$ where $n$ is the dimension of $\mathbf x$.
Potentials are functions of a three dimensional position variable.  Following \citet{DZ05}, we use $\tilde\phi$ to denote the Fourier transform of the potential in the angular (transverse) variables only
\be
\tilde\phi\left(\mathbf l; \chi\right) \equiv\frac{1}{\chi^2}\int \frac{\mr d k_3}{2\pi}\tilde\Phi(\mathbf l/\chi, k_3;z(\chi))\mr e ^{\mr i k_3 \chi}\;.
\ee  
Then the spatial derivatives of the potential can be expressed in terms of the angular Fourier transform $\tilde\phi$ as
\be
\Phi_{,i_1i_2...i_M}(\mathbf{n}_0\chi;\chi) = \frac{\mr i^M}{\chi^M} \int \frac{\mr d^2\mathbf l}{(2\pi)^2}
l_{i_1}l_{i_2}...l_{i_M}
\tilde\phi({\mathbf l}; \chi) \mr e^{\mr i\mathbf l\cdot\mathbf n_0}.
\ee
Applying this to the first term from Eq. (\ref{eq:shear}) and using the relation between convergence, shear and $\psi_{ij}$, we arrive at the well-known first order results for convergence and shear
\be
\tilde{\kappa}^{(1)}(\mathbf l, z_{\mr s})  = \frac{1}{2}\left(\tilde\psi^{(1)}_{11}(\mathbf l, z_{\mr s})+\tilde\psi^{(1)}_{22}(\mathbf l, z_{\mr s}) \right)=-l^2 \int_0^{\chi_{\mr s}}\!\!\! \mr d \chi W(\chi, \chi_{\mr s})\tilde{\phi}({\mathbf l}; \chi)
{\rm ~~and~~}
\tilde{\gamma}^{(1)}_I(\mathbf l, z_{\mr s}) = T_I(\mathbf l) \tilde{\kappa}^{(1)}(\mathbf l, z_{\mr s})\;.
\label{eq:firstorder}
\ee
Here $T_1(\mathbf l) = \cos(2\phi_l)$ and $T_2(\mathbf l)=  \sin(2\phi_l)$, where $\phi_l$ is the azimuthal angle of $\mathbf l$.

We generally decompose the shear components into tangential (or E-mode) shear, $\gamma_{\mr E}$ and cross (or B-mode) shear, $\gamma_{\mr B}$,
\be
\tilde{\gamma}_{\mr E} (\mathbf l, z_{\mr s})=  \delta_{IJ} T_{I}(\mathbf{l}) \tilde{\gamma}_J(\mathbf l, z_{\mr s});
{\rm ~~~~}
 \tilde{\gamma}_{\mr B} (\mathbf l, z_{\mr s})   
=   \epsilon_{IJ} T_{I}(\mathbf l)\tilde{\gamma}_J(\mathbf l, z_{\mr s})\;,
\label{eq:t}
\ee
with $\epsilon_{IJ}$ the two dimensional Levi-Civita tensor. To first order, $\tilde{\gamma}^{(1)}_{\mr E} (\mathbf l, z_{\mr s}) = \tilde{\kappa}^{(1)}(\mathbf l, z_{\mr s})$ and $ \tilde{\gamma}_{\mr B}^{(1)} (\mathbf l, 
z_{\mr s})=\tilde{\omega}^{(1)} (\mathbf l, z_{\mr s})= 0$.
Their power spectra can be obtained under the Limber approximation \citep[ Eq. (15)]{Kaiser92,DZ05},
\be
\langle \tilde\phi({\mathbf l};\chi) \tilde\phi({\mathbf l'};\chi') \rangle = (2\pi)^2 \delta_{\mr D}(\mathbf l+\mathbf l')
\frac{\delta_{\mr D}(\chi-\chi')}{\chi^2} P_{\Phi}\left( l/\chi;z(\chi)\right),
\label{eq:limber2}
\ee
where $P_{\Phi}\left(l/\chi;z(\chi)\right)$ is the three dimensional power spectrum of the potential at redshift $z(\chi)$. 
The lensing tomography cross spectra between two source redshift slices at $z_\alpha$ and $z_\beta$ (with $z_\alpha<z_\beta$) then read
\be
C^{(11)}_{\tilde{\gamma}_{\mr E}}(l;z_\alpha,z_\beta) =C^{(11)}_{\tilde{\kappa}}(l;z_\alpha,z_\beta) = l^4\int_0^{\chi_\alpha}\mr d \chi \frac{W(\chi,\chi_\alpha)W(\chi,\chi_\beta)}{\chi^2}P_{\Phi}\left(l/\chi;z(\chi)\right)
\ee
and
\be
C^{(11)}_{ \tilde{\gamma}_{\mr B}}(l;z_\alpha,z_\beta) =  C^{(11)}_{\tilde{\omega}}(l;z_\alpha,z_\beta) = 0\;,
\ee
where the superscripts denote the order of expansion in the potential.

\subsection{Fourier space: second order}
To work to second order, we need the usual convolution theorem for the product of two fields $U$ and $V$ is
\be
\widetilde{[UV]}(\mathbf l) \equiv [\tilde U\ast\tilde V](\mathbf l) = \int \frac{\mr d^2\mathbf l'}{(2\pi)^2} \tilde U(\mathbf l')\tilde V(\mathbf l-\mathbf l')\;.
\ee
Introducing 
\be
\mathfrak{M}\left(\mathbf{l'},\mathbf{l-l'}; z_{\mr s}\right)  = \int_0^{\chi_{\mr s}}\!\!\! \mr d \chi W(\chi, \chi_{\mr s})\!\int_0^{\chi}\!\!\! \mr d \chi' 
W(\chi', \chi) \tilde{\phi}\left(\mathbf l'; \chi\right)\tilde{\phi}\left(\mathbf{l-l'}; \chi'\right),
\label{eq:scriptM}
\ee
and using the second term from Eq. (\ref{eq:shear}) and the relation between convergence, rotation, shear and $\psi_{ij}$, the second order corrections to convergence, rotation and shear can be written as
\be
\tilde{\kappa}^{(2)}(\mathbf l, z_{\mr s}) =
-2\int\!\! \frac{\mr{d}^2\mathbf l'}{(2\pi)^2} \left[ \mathbf{l'}\cdot\left(\mathbf l - \mathbf{l'}\right)\right] \mathbf l\cdot \mathbf{l'}
 \mathfrak{M}\left(\mathbf{l'},\mathbf{l-l'}; z_{\mr s}\right)\;,
 {\rm ~~~~}
\tilde{\omega}^{(2)}(\mathbf l, z_{\mr s}) =
-2\int\!\! \frac{\mr{d}^2\mathbf l'}{(2\pi)^2} \left[ \mathbf{l'}\cdot\left(\mathbf l - \mathbf{l'}\right)\right]  ll' \sin(\phi_{l'})
 \mathfrak{M}\left(\mathbf{l'},\mathbf{l-l'}; z_{\mr s}\right)
\label{eq:kappa2}
\ee
and
\be
\tilde{\gamma}_I^{(2)}(\mathbf l, z_{\mr s}) =
-2\int\!\! \frac{\mr{d}^2\mathbf l'}{(2\pi)^2} \left[ \mathbf{l'}\cdot\left(\mathbf l - \mathbf{l'}\right)\right] l l' G_I(\mathbf l,\mathbf{l'})
 \mathfrak{M}\left(\mathbf{l'},\mathbf{l-l'}; z_{\mr s}\right).
\label{eq:gamma2}
\ee
Here the superscript refers to the order of expansion in $\Phi$, and we define
$G_1(\mathbf l,\mathbf{l'}) =\cos(\phi_l + \phi_{l'})$ and
$G_2(\mathbf l,\mathbf{l'}) =\sin(\phi_l + \phi_{l'})$.
When we work beyond first order in the lensing potential, the shear becomes a non-linear function of the gravitational potential $\Phi$. Hence the power spectrum of 
the shear depends on the higher order correlation functions of $\Phi$.  Therefore we need the Limber approximation for these higher order correlation functions.  
For the bispectrum, Eq.~(\ref{eq:limber2}) generalizes to
\be
\ensav{ \tilde\phi\left(\mathbf l_1;\chi_1\right) \tilde\phi\left(\mathbf l_2;\chi_2\right) \tilde\phi\left(\mathbf l_3;\chi_3\right) }
= (2\pi)^2 \delta_{\mr D}(\mathbf l_1+\mathbf l_2+\mathbf l_3)
\frac{
\delta_{\mr D}(\chi_1-\chi_2)
\delta_{\mr D}(\chi_1-\chi_3)
}{\chi_1^4}
B_{\Phi}\left(\frac{\mathbf l_1}{\chi_1},\frac{\mathbf l_2}{\chi_1}, \frac{\mathbf l_3}{\chi_1};z(\chi_1)\right);
\label{eq:limber3}
\ee
and for the trispectrum,
\be
\ensav{  \tilde\phi\left(\mathbf l_1;\chi_1\right)\tilde\phi\left(\mathbf l_2;\chi_2\right)\tilde\phi\left(\mathbf l_3;\chi_3\right)\tilde\phi\left(\mathbf l_4;\chi_4\right)}_{\mr c}
= (2\pi)^2 \delta_{\mr D}(\mathbf l_1+\mathbf l_2+\mathbf l_3+\mathbf l_4)
\frac{\delta_{\mr D}(\chi_1-\chi_2)
\delta_{\mr D}(\chi_1-\chi_3)
\delta_{\mr D}(\chi_1-\chi_4)}{\chi_1^6}
T_{\Phi}\left(\frac{\mathbf l_1}{\chi_1}, \frac{\mathbf l_2}{\chi_1},\frac{\mathbf l_3}{\chi_1}, \frac{\mathbf l_4}{\chi_1};z(\chi_1) \right),
\label{eq:limber4}
\ee
where the subscript ``c'' denotes a connected function.

As an example, we consider the correlation of two $\mathfrak{M}$ functions,
\beq
\ensav{\mathfrak{M}\left(\mathbf{l'},\mathbf{l-l'}; z_\alpha \right)\mathfrak{M}\left(\mathbf{l'''},\mathbf{l''-l'''}; z_\beta \right)}& =&
\int_0^{\chi_\alpha} \mr d\chi \int_0^{\chi_\beta}\mr d\chi'' \int_0^{\chi} \mr d\chi' \int_0^{\chi''}\mr d\chi'''
W(\chi,\chi_\alpha) W(\chi',\chi) W(\chi'',\chi_\beta) W(\chi''',\chi'')
\nonumber \\& & \times
\ensav{ \tilde\phi\left(\mathbf l';\chi\right) \tilde\phi\left(\mathbf l-\mathbf l';\chi'\right) \tilde\phi\left(\mathbf l''';\chi''\right) \tilde\phi\left(\mathbf l''-\mathbf l''';\chi'''\right)}.
\eeq
The expectation value here can be broken up into a Gaussian (Wick's theorem) piece and a connected (non-Gaussian) piece. The connected piece vanishes because the 
$\delta_{\mr D}$-functions in Eq.~(\ref{eq:limber4}) force $\chi=\chi'=\chi''=\chi'''$ where the window functions vanish. Of the 3 possible contractions for the 
Gaussian term, the only one that survives is $\chi''=\chi>\chi'''=\chi'$. Thus,
\be
\ensav{\mathfrak{M}\left(\mathbf{l'},\mathbf{l-l'}; z_\alpha \right)\mathfrak{M}\left(\mathbf{l'''},\mathbf{l''-l'''}; z_\beta \right)} 
= (2\pi)^4\delta_{\mr D}(\bfmath l'+\bfmath l''')\delta_{\mr D}(\bfmath l -\bfmath l'+\bfmath l''-\bfmath l''' ) M\left(l',|\mathbf 
l-\mathbf l'|;z_\alpha,z_\beta\right),
\label{eq:MPS}
\ee
where we have introduced the mode-coupling integral
\be
M\left(l, l';z_\alpha,z_\beta \right) =  \int_0^{\chi_\alpha} \!\!\mr d \chi \frac{W(\chi,\chi_\alpha)W(\chi,\chi_\beta)}{\chi^2}\!\!\! \int_0^{\chi}\!\! \mr d \chi' \frac{W^2(\chi',\chi)}{\chi'^2}P_{\Phi}\left(l/\chi;z(\chi)\!\right)\!P_{\Phi}\left(l'/\chi';z(\chi')\!\right)\!\, .
\ee
Note that Eq.~(\ref{eq:MPS}) is true even for a non-Gaussian density field.

The third order terms each require specialized treatment, so we handle them on a case-by-case basis below.

\section{The corrections to the power spectrum}
\label{sec:corrections}
We can now calculate the higher order contributions to the reduced shear power spectrum by Taylor expanding the reduced shear in terms of the shear and convergence to contain all terms up to $\mathcal O (\Phi^4)$,
\be
\ensav{\tilde{g}_{\mr E/\mr B}(\mathbf l, z_\alpha)\;\tilde{g}_{\mr E/\mr B}(\mathbf l', z_\beta)} \approx  \ensav{(\tilde{\gamma}\ast(1+\tilde{\kappa}+ \tilde{\kappa}\ast\tilde{\kappa}))_{\mr E/\mr B}(\mathbf l, z_\alpha)\;(\tilde{\gamma}\ast(1+\tilde{\kappa}+\tilde{\kappa}\ast\tilde{\kappa}))_{\mr E/\mr B}(\mathbf l', z_\beta)}\; , 
\label{eq:gg}
 \ee
where $\ast$ denotes a convolution, and where the shear and convergence need to be expanded in terms of the potential according to Eq. (\ref{eq:shear}) and projected into $\mr{E/B}$ components using Eq. (\ref{eq:t}).

As the power spectra depend only on the magnitude of $\mathbf l$, we can choose $\mathbf l \| \hat{\mathbf x}$, which implies $T(\mathbf l) = (1,0)$ and thus $\tilde\gamma_{\mr E}(\mathbf l) = \tilde\gamma_1(\mathbf l)$,  and simplifies the calculations without loss of generality. Consider for example the correction to the E-mode power spectrum arising from the correlation between second order corrections
\begin{align}
\nonumber \ensav{\tilde{g}^{(2)}_{\mr E}(\mathbf l, z_\alpha)\;\tilde{g}^{(2)}_{\mr E}(\mathbf l', z_\beta)} = &\;\ensav{\left(\tilde{\gamma}^{(2)}+\tilde{\gamma}^{(1)}\!\ast\tilde{\kappa}^{(1)}\right)_{\mr E}(\mathbf l, z_\alpha)\;\left(\tilde{\gamma}^{(2)}+\tilde{\gamma}^{(1)}\!\ast\tilde{\kappa}^{(1)}\right)_{\mr E}(\mathbf l', z_\beta)}\\
\nonumber =&\; \ensav{\tilde\gamma^{(2)}_1(\mathbf l, z_\alpha)\;\tilde\gamma^{(2)}_{\mr E}(\mathbf l', z_\beta)}+ \ensav{\left(\tilde{\gamma}^{(1)}_1\!\ast\tilde{\kappa}^{(1)}\right)(\mathbf l, z_\alpha)\;\left(\tilde{\gamma}^{(1)}\!\ast\tilde{\kappa}^{(1)}\right)_{\mr E}(\mathbf l', z_\beta)} \\
\nonumber &+\ensav{\tilde\gamma^{(2)}_1(\mathbf l, z_\alpha)\;\left(\tilde{\gamma}^{(1)}\!\ast\tilde{\kappa}^{(1)}\right)_{\mr E}(\mathbf l', z_\beta)} + \ensav{\left(\tilde{\gamma}^{(1)}_1\!\ast\tilde{\kappa}^{(1)}\right)(\mathbf l, z_\alpha)\;\gamma^{(2)}_{\mr E}(\mathbf l', z_\beta)}\\
=&\;  \ensav{\tilde{\gamma}_1^{(2)}(\mathbf{l})\;T_I(\mathbf{l'}) \tilde{\gamma}_I^{(2)}(\mathbf{l'})}_{\alpha \beta} +\ensav{\left(\tilde{\gamma}_1^{(1)}\!\ast\tilde{\kappa}^{(1)}\right)\!(\mathbf{l})\;T_I(\mathbf{l'})\left(\tilde{\gamma}_I^{(1)}\!\ast\tilde{\kappa}^{(1)}\right)\!(\mathbf{l'})}_{\alpha \beta}
+ 2\ensav{\tilde{\gamma}_1^{(2)}(\mathbf{l})\;T_I(\mathbf{l'})\left(\tilde{\gamma}_I^{(1)}\!\ast\tilde{\kappa}^{(1)}\right)\!(\mathbf{l'})}_{\alpha \beta}\;,
\end{align}
where in the last step we have rewritten the E-mode component using Eq. (\ref{eq:t}) and where we define the symmetrized expectation value
\be
\ensav{A(\mathbf l)B(\mathbf l')}_{\alpha\beta} = \frac12 \left[\ensav{A(\mathbf l, z_\alpha)B(\mathbf l',z_\beta)}+\ensav{A(\mathbf l, z_\beta)B(\mathbf 
l',z_\alpha)}\right],
\ee
to shorten our notation.

Noting $\tilde{\gamma}_{\mr B}^{(1)}(\mathbf l) = 0$ and $\ensav{\tilde{\gamma}_{\mr E} (\mathbf l)\tilde{\gamma}_{\mr B} (\mathbf l')} = 0$, we can expand Eq. (\ref{eq:gg}) to $\mathcal O (\Phi ^4)$:
\begin{align}
\nonumber \ensav{\tilde{g}_{\mr E}(\mathbf l, z_\alpha)\;\tilde{g}_{\mr E}(\mathbf l', z_\beta)} =&\;  C^{(11)}_{\tilde\gamma_{\mr E}}(l;z_\alpha,z_\beta) +\underbrace{\Delta C^{(12)}_{\tilde{g}_{\mr E}}(l;z_\alpha,z_\beta) }_{\mathcal O(\Phi^3) \;\mr{reduced\; shear}}+ \underbrace{\Delta C^{(13)}_{\tilde\gamma_{\mr E}}(l;z_\alpha,z_\beta)  +\Delta C^{(22)}_{\tilde\gamma_{\mr E}}(l;z_\alpha,z_\beta) }_{\mathcal O(\Phi^4) \;\mr{ shear}}+ \underbrace{\Delta C^{(13)}_{\tilde{g}_{\mr E}}(l;z_\alpha,z_\beta)  +\Delta C^{(22)}_{\tilde{g}_{\mr E}}(l;z_\alpha,z_\beta) }_{\mathcal O(\Phi^4) \;\mr{reduced\; shear}}  \\
\nonumber =&\; C^{(11)}_{\tilde\gamma_{\mr E}}(l;z_\alpha,z_\beta) 
+ 2\ensav{\tilde{\gamma}_1^{(1)}(\mathbf{l})\;T_I(\mathbf{l'})\left(\tilde{\gamma}_I^{(1)}\!\ast\tilde{\kappa}^{(1)}\right)\!(\mathbf{l'})}_{\alpha \beta}
\! + 2\ensav{\tilde{\gamma}_1^{(1)}(\mathbf{l}) \;T_I(\mathbf{l'})\tilde{\gamma}_I^{(3)}(\mathbf{l'})}_{\alpha \beta}  
\!+  \ensav{\tilde{\gamma}_1^{(2)}(\mathbf{l})\;T_I(\mathbf{l'}) \tilde{\gamma}_I^{(2)}(\mathbf{l'})}_{\alpha \beta}
   \\
\nonumber
 &+2\left\{\ensav{\tilde{\gamma}_1^{(1)}(\mathbf{l})\;T_I(\mathbf{l'})\left(\tilde{\gamma}_I^{(1)}\!\ast\tilde{\kappa}^{(2)}\right)\!(\mathbf{l'})}_{\alpha \beta}
\!+\ensav{\tilde{\gamma}_1^{(1)}(\mathbf{l})\;T_I(\mathbf{l'})\left(\tilde{\gamma}_I^{(2)}\!\ast\tilde{\kappa}^{(1)}\right)\!(\mathbf{l'})}_{\alpha \beta} 
\!+\ensav{\tilde{\gamma}_1^{(1)}(\mathbf{l})\;T_I(\mathbf{l'})\left(\tilde{\gamma}_I^{(1)}\!\ast\tilde{\kappa}^{(1)}\!\ast\tilde{\kappa}^{(1)}\right)\!(\mathbf{l'})}_{\alpha \beta}\right\}\\
\label{eq:gBig}
 &+\left\{2\ensav{\tilde{\gamma}_1^{(2)}(\mathbf{l})\;T_I(\mathbf{l'})\left(\tilde{\gamma}_I^{(1)}\!\ast\tilde{\kappa}^{(1)}\right)\!(\mathbf{l'})}_{\alpha \beta}
\!
+\ensav{\left(\tilde{\gamma}_1^{(1)}\!\ast\tilde{\kappa}^{(1)}\right)\!(\mathbf{l})\;T_I(\mathbf{l'})\left(\tilde{\gamma}_I^{(1)}\!\ast\tilde{\kappa}^{(1)}\right)\!(\mathbf{l'})}_{\alpha \beta}\right\}\\
\nonumber \ensav{\tilde{g}_{\mr B}(\mathbf l, z_\alpha)\;\tilde{g}_{\mr B}(\mathbf l', z_\beta)} = &
\;\Delta C^{(22)}_{\tilde\gamma_{\mr B}}(l;z_\alpha,z_\beta) +\Delta C^{(22)}_{\tilde{g}_{\mr B}}(l;z_\alpha,z_\beta)\\ 
=& 
\ensav{\tilde{\gamma}_2^{(2)}(\mathbf{l}) \;\epsilon_{IJ}T_I(\mathbf{l'})\tilde{\gamma}_J^{(2)}(\mathbf{l'})}_{\alpha \beta}
\!+2\ensav{\tilde{\gamma}_2^{(2)}(\mathbf{l}) \;\epsilon_{IJ}T_I(\mathbf{l'})\left(\tilde{\gamma}_J^{(1)}\!\ast\tilde{\kappa}^{(1)}\right)\!\!(\mathbf{l'})}_{\alpha \beta}
\!+\ensav{\left(\tilde{\gamma}_2^{(1)}\!\ast\tilde{\kappa}^{(1)}\right)\!\!(\mathbf{l}) \;\epsilon_{IJ}T_I(\mathbf{l'})\left(\tilde{\gamma}_J^{(1)}\!\ast\tilde{\kappa}^{(1)}\right)\!\!(\mathbf{l'})}_{\alpha \beta},
\end{align}
where we have omitted terms such as $\Delta C^{(12)}_{\tilde\gamma_{\mr E}}$ which vanish under the Limber approximation.

\subsection{Multiple-deflection shear corrections}
\label{sec:shear}
The shear-only corrections come in two flavors: the ``22'' (2nd order-2nd order) terms and the ``13'' terms.  The ``12'' terms are mathematically of order $\Phi^3$, 
and hence one might expect them to be present if the matter bispectrum is non-zero.  However, they vanish in the Limber approximation due to the $W(\chi',\chi)$ factor in 
Eq.~(\ref{eq:scriptM}), which is zero whenever $\chi'=\chi$.

The ``22'' B-mode shear correction can be written as
\begin{align}
\nonumber \ensav{\tilde{\gamma}_{\mr B}^{(2)}(\mathbf{l}, z_\alpha) \tilde{\gamma}_{\mr B}^{(2)}(\mathbf{l''},z_\beta)} = &\;
2\epsilon_{IJ}T_I(\mathbf l) \int\!\! \frac{\mr{d}^2l'}{(2\pi)^2} \left[ \mathbf{l'}\cdot\left(\mathbf l - \mathbf{l'}\right)\right] l l' G_J(\mathbf l,\mathbf{l'})\;2\epsilon_{HK}T_H(\mathbf l'') \\
\nonumber &\times  \int\!\! \frac{\mr{d}^2l'''}{(2\pi)^2} \left[ \mathbf{l}'''\cdot\left(\mathbf l'' - \mathbf{l'''}\right)\right] l'' l''' G_K(\mathbf l'',\mathbf{l'''}) 
 \ensav{\mathfrak{M}\left(\mathbf{l'},\mathbf{l-l'}; z_\alpha \right)\mathfrak{M}\left(\mathbf{l'''},\mathbf{l''-l'''}; z_\beta \right)}\\
 = &\;(2\pi)^2 \delta(\bfmath l+\bfmath l'')
 4l^2\int\!\! \frac{\mr{d}^2l'}{(2\pi)^2} \left(l'\sin\phi_{l'}\right)^2 \left[\mathbf{l'}\cdot\left(\mathbf l - \mathbf{l'}\right)\right]^2 
 M\left(l',|\mathbf l-\mathbf l'|;z_\alpha,z_\beta \right),
 \label{eq:gammacross22}
\end{align}
where we have used Eqs. (\ref{eq:t}, \ref{eq:gamma2}, \ref{eq:MPS}) and $\phi_l = 0$ repeatedly. By comparison with Eq. (\ref{eq:kappa2}) one can see that $\Delta C^{(22)}_{\tilde\gamma_{\mr B}} = \Delta C^{(22)}_{\tilde\omega}$. Similarly,
\be
\Delta C^{(22)}_{\tilde\gamma_{\mr E}}(l; z_\alpha,z_\beta) = 4l^2\int\!\! \frac{\mr{d}^2\mathbf l'}{(2\pi)^2} \left(l'\cos\phi_{l'}\right)^2
 \left[\mathbf{l'}\cdot\left(\mathbf l - \mathbf{l'}\right)\right]^2 
 M\left(l',|\mathbf l-\mathbf l'|;z_\alpha,z_\beta \right),
 \label{eq:gamma22}
\ee
and 
\be
\Delta C^{(22)}_{\tilde\kappa}(l; z_\alpha,z_\beta) = 4\int\!\! \frac{\mr{d}^2\mathbf l'}{(2\pi)^2} \left(\mathbf l \cdot \mathbf l'\right)^2
 \left[\mathbf{l'}\cdot\left(\mathbf l - \mathbf{l'}\right)\right]^2 
 M\left(l',|\mathbf l-\mathbf l'|;z_\alpha,z_\beta \right)\;.
\label{eq:kappa22}
\ee
The integrals in Eqs.~(\ref{eq:gamma22}, \ref{eq:kappa22}) are dominated by angular scales corresponding to the peak of the matter power spectrum, which is at scales much larger than those typically probed by lensing:
If we define $\mathbf l_{\mr c} = \mathbf l - \mathbf l'$, then for small $l_{\mr c}$ (compared to $l$ of lensing experiments) the contribution to these integrals scales as $\int \mr{d}^2 \mathbf l_{\mr c} l_{\mr c}^2 \cos^2(\mathbf l, \mathbf l_{\mr c}) M(l,l_{\mr c};z_\alpha,z_\beta)$. 
Assuming an effective power-law index $n_{\mr s}^{\mr{eff}}$ for the non-linear matter power spectrum $P_{\delta,\mr nl}(k)$, the $l_{\mr c}$-dependence of $M(l,l_{\mr c};z_\alpha,z_\beta)$ scales as $l_c^{n_{\mr s}^{\mr{eff}}-4}$. 
So the contribution to the integral per logarithmic range in $l_{\mr c}$ scales as $l_{\mr c}^{n_{\mr s}^{\mr{eff}}}$, which is dominated by scales corresponding to the peak of the matter power spectrum.

The ``13'' correction in principle has three parts: those arising from the 3A, 3B, and 3C terms of Eq.~(\ref{eq:shear}).  Let us consider the 3B term first.  The 
expectation value of the product of two Fourier modes is
\beq
\langle \psi_{ab}^{(1)}(\mathbf l,z_\alpha) \psi_{ij}^{(3B)}(\mathbf L,z_\beta)\rangle &=&
16 \int_0^{\chi_\alpha} \mr d\chi \int_0^{\chi_\beta} \mr d\chi_1 \int_0^{\chi_1} \mr d\chi'_1 \int_0^{\chi_1} \mr d\chi''_1 
\int \frac{\mr d^2\mathbf L'}{(2\pi)^2} \int \frac{\mr d^2\mathbf L''}{(2\pi)^2}
W(\chi,\chi_\alpha) W(\chi_1,\chi_\beta) W(\chi_1',\chi_1)W(\chi''_1,\chi_1)
\nonumber \\ && \times
l_al_b
L'_cL'_jL''_d
(\mathbf L-\mathbf L'-\mathbf L'')_i (\mathbf L-\mathbf L'-\mathbf L'')_c (\mathbf L-\mathbf L'-\mathbf L'')_d
\nonumber \\ && \times
\langle \tilde\phi(\mathbf l;\chi)
\tilde\phi(\mathbf L-\mathbf L'-\mathbf L'';\chi_1)
\tilde\phi(\mathbf L';\chi'_1) \tilde\phi(\mathbf L'';\chi''_1)\rangle\;.
\eeq
In the Limber approximation, the only non-vanishing contraction is at $\chi=\chi_1$ and $\chi'_1=\chi''_1$.  The $\delta_{\mr D}$-functions then enforce $\mathbf 
L'_1=-\mathbf L''_1$ and $\mathbf L=-\mathbf l$.  We thus find:
\beq
\langle \psi_{ab}^{(1)}(\mathbf l,z_\alpha) \psi_{ij}^{(3B)}(\mathbf L,z_\beta)\rangle &=&
(2\pi)^2\delta_{\mr D}(\mathbf l+\mathbf L)
16 \int_0^{\min(\chi_\alpha,\chi_\beta)} \mr d\chi \int_0^\chi \mr d\chi'_1
\frac{W(\chi,\chi_\alpha)W(\chi,\chi_\beta)W^2(\chi_1',\chi)}{\chi^2\chi_1'{^2}}
\nonumber \\ && \times
\int \frac{\mr d^2\mathbf L'}{(2\pi)^2} P_\Phi\left( l/\chi;z(\chi)\right) P_\Phi\left(L'/\chi'_1;z(\chi'_1)\right)
l_al_bL'_cL'_jL'_dl_il_cl_d.
\label{eq:13b}
\eeq
The integrand is odd under $\mathbf L'\rightarrow -\mathbf L'$, and hence the ``13B" correction to the shear power spectrum vanishes.

The ``13C'' correction is zero because the restriction $\chi''<\chi'<\chi$ in Eq.~(\ref{eq:shear}) implies that there are no allowed contractions within the 
independent 
lens 
plane approximation.  This leaves us with the ``13A" correction, which is similar to ``13B", except with the replacement $L'_j\rightarrow l_j$. The choice $\mathbf l || \hat{\mathbf x}$ implies that the only non-vanishing component of "13A" is  $\langle \psi_{11}^{(1)}(\mathbf l,z_\alpha) \psi_{11}^{(3A)}(\mathbf L,z_\beta)\rangle$. Hence we find
\be
\Delta C^{(13)}_{\tilde\gamma_{\mr E}}(l; z_\alpha,z_\beta) = 
\Delta C^{(13A)}_{\tilde\gamma_{\mr E}}(l; z_\alpha,z_\beta) = \Delta C^{(13A)}_{\tilde\kappa}(l; z_\alpha,z_\beta) = 
-4l^4\int\!\!\frac{\mr d^2\mathbf l'}{(2\pi)^2}\left(\mathbf{l}\cdot \mathbf{l'}\right)^2 
M(l,l';z_\alpha,z_\beta)\;.
 \label{eq:gamma13}
\ee
There is no ``13" B-mode shear or rotation power spectrum because $\tilde\gamma^{(1)}_{\mr B}(\mathbf l,z_\alpha)$ and $\tilde\omega^{(1)}(\mathbf l,z_\alpha)$ vanish.

The dimensionless shear power spectrum, $\Delta^{2_{(11)}}_{\tilde\gamma_{\mr E}}(l)=l(l+1) C^{(11)}_{\tilde\gamma_{\mr E}}(l)/(2\pi)^2$ 
scales as $\Delta^{2_{(11)}}_{\tilde\gamma_{\mr E}}(l) \propto l^{n_{\mr s}^{\mr{eff}}+2}$, while the corrections $\Delta^{2_{(13)}}_{\tilde\gamma_{\mr E}}(l)$ and 
$\Delta^{2_{(22)}}_{\tilde\gamma_{\mr E}}(l)$ scale as $l^{n_{\mr s}^{\mr{eff}}+4}$. The main contribution to these corrections at large $l$ is the bulk deflection on small 
scales by large wavelength density perturbations which causes only small local distortions. Thus the ``22'' and ``13'' terms largely cancel, similar to the 
perturbative 
calculation of the one-loop correction to the density power spectrum \citep[e.g.][]{V83}.
As these corrections diverge for large $l$ and have opposite sign, their numerical difference needs to be evaluated carefully\footnote{Apply a variable transform $\mathbf l'' = \mathbf l - \mathbf l'$ to $\Delta C^{(22)}_{\tilde\gamma_{\mr E}}$ and cancel diverging contributions at $\mathbf l''$ by rewriting the integral  as $\Delta C^{(22)}_{\tilde\gamma_{\mr E}}+\Delta C^{(13)}_{\tilde\gamma_{\mr E}} = 4\int\!\!\frac{\mr d^2\mathbf l''}{(2\pi)^2}\left(\mathbf{l}\cdot \left(\mathbf{l''}+\mathbf l\right)\right)^2 \left(\mathbf{l''}\cdot \left(\mathbf{l''}+\mathbf l\right)\right)^2 \left( M(|\mathbf l+\mathbf l''|, l'';z_\alpha, z_\beta) - M(l,l'';z_\alpha,z_\beta)\right) + \int\!\!\frac{\mr d^2\mathbf l''}{(2\pi)^2}\left(\left(\mathbf{l}\cdot \left(\mathbf{l''}+\mathbf l\right)\right)^2 \left(\mathbf{l''}\cdot \left(\mathbf{l''}+\mathbf l\right)\right)^2 -l^4\left(\mathbf l \cdot \mathbf l ''\right)^2\right)M(l,l'';z_\alpha,z_\beta) $, where the azimuthal integration of the second term can be done analytically.}.

\begin{figure}
\includegraphics[width = 0.5\textwidth]{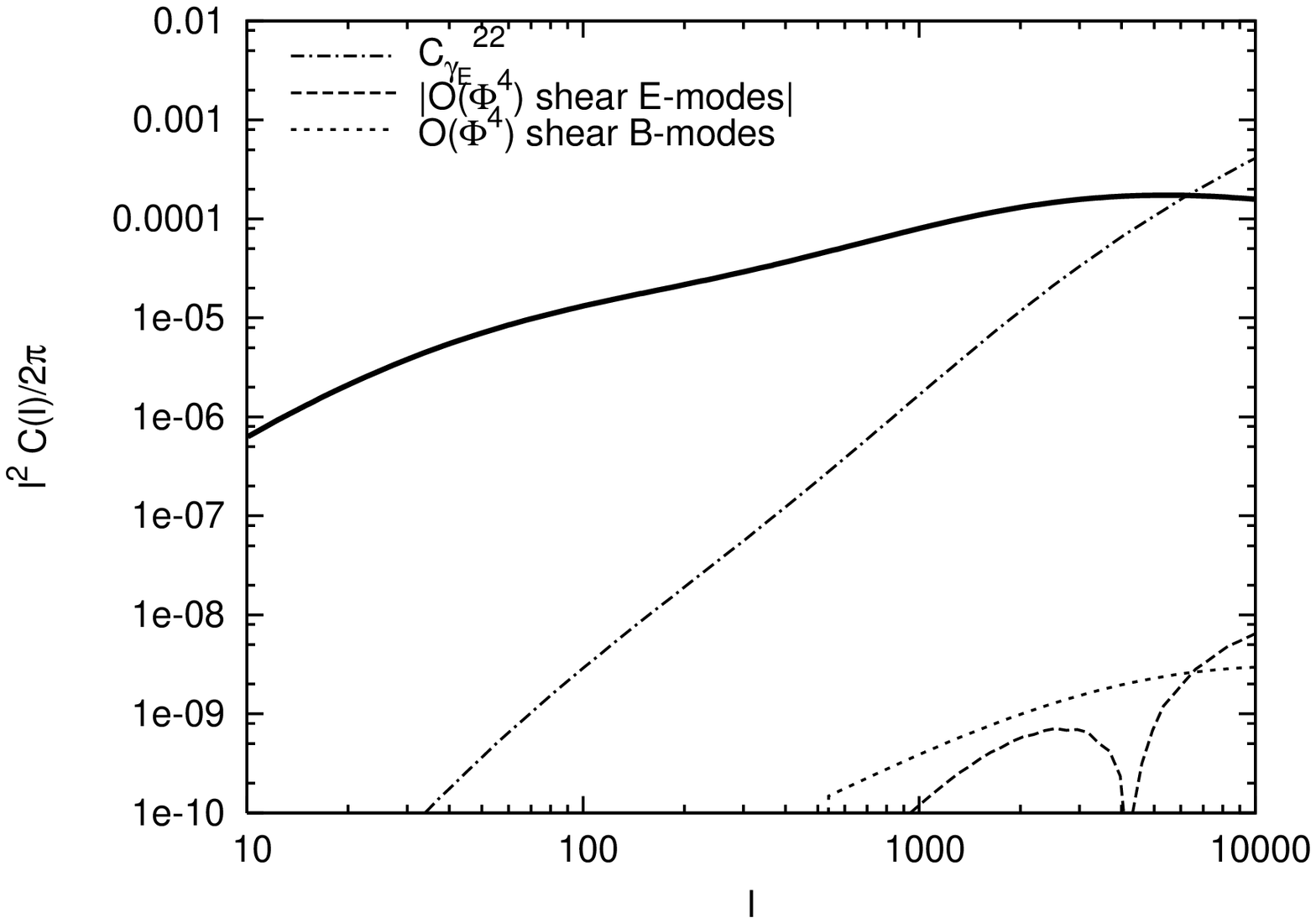}
\includegraphics[width = 0.5\textwidth]{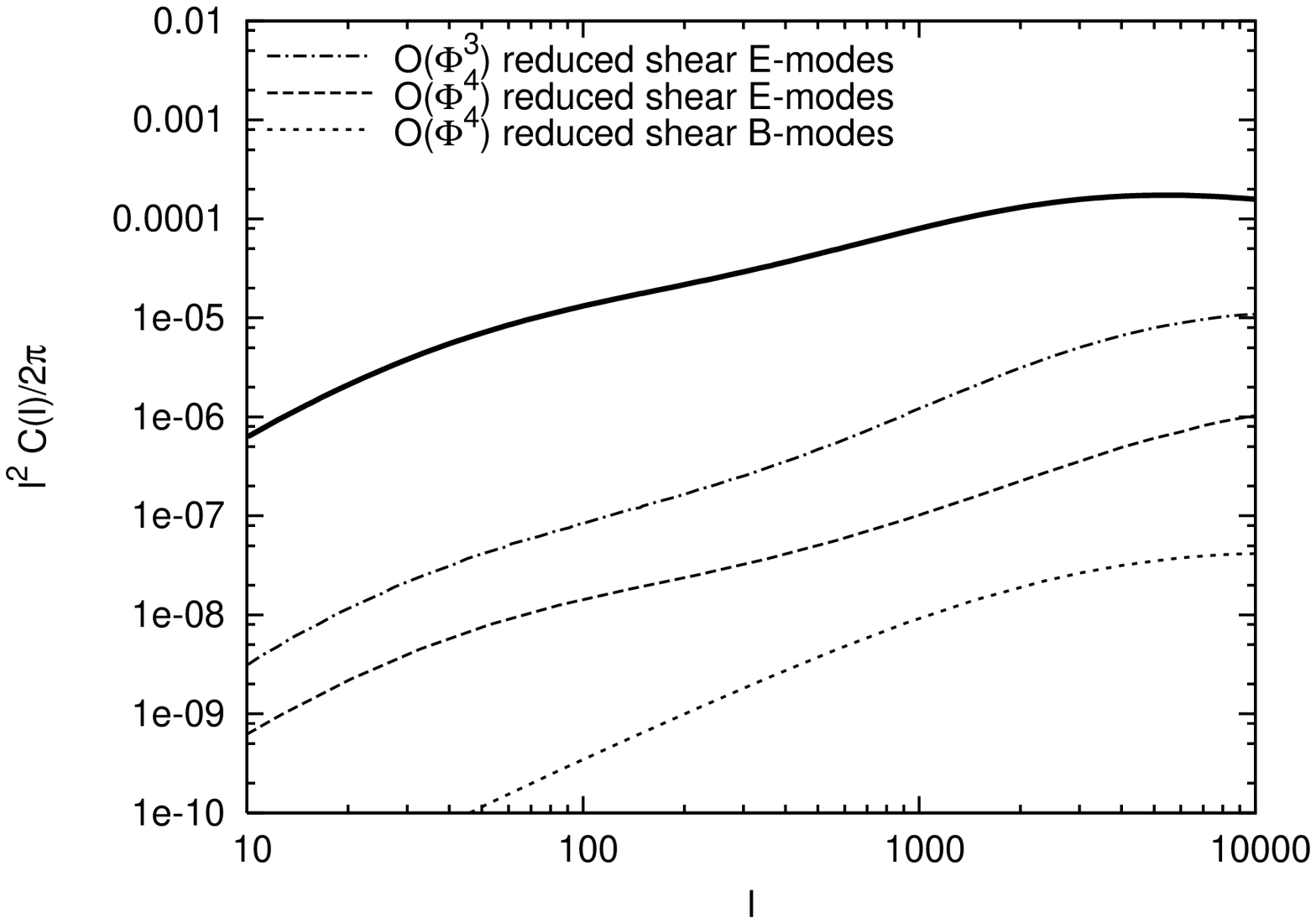}
\caption{Linear order shear power spectrum (thick solid line; Eq. (\ref{eq:firstorder})) and corrections up to $\mathcal{O}(\Phi^4)$.\newline
\emph{Left.} The dashed/short dashed lines show the fourth order corrections to the E/B-mode shear power spectra that arise from relaxing the Born approximation and including lens-lens coupling in the calculation of the shear (Sect.~\ref{sec:shear}; cf. \citet{CH}). The E-mode correction is negative at small $l$ and positive for $l\gtrsim 4200$. The dashed-dotted line illustrates term $C_{\gamma_{\mr E}}^{(22)}$ (cf. Eq.~(\ref{eq:gamma22})) which contributes to the E-mode shear correction , the divergency is cancelled by Eq.~(\ref{eq:gamma13}).\newline
\emph{Right.} The dashed/short dashed lines show the combined forth order corrections to the reduced shear E/B-mode power spectra (Sect.~\ref{sec:cg}, Table~\ref{tab:cg}). The dashed-dotted line shows the third order correction to the reduced shear E-mode power spectrum.\newline
We assume a source redshift $z_{\alpha}=z_\beta= 1$ and use the transfer function from \citet{ebw92}, the fitting formula of \citet{sm03} for the non-linear matter power spectrum, and the fitting formula of \citet{SC_bispect} for the non-linear matter bispectrum.
This figure assumes a flat $\Lambda$CDM cosmology with $(\Omega_{\mr m},\Omega_{\mr b},\sigma_8, h, n) = (0.3, 0.05,0.9, 0.7,1)$ to enable comparison with previous calculations.}
\label{fig:Cg}
\end{figure}

The dotted lines in Fig. \ref{fig:Cg} illustrate their magnitude for $z_\alpha = z_\beta =1$ using the fitting formula of \citet{sm03}
for the non-linear matter power spectrum with the transfer function from \citet{ebw92} for the numerical integration.
Here the combined E-mode correction is negative at small $l$
and positive for $l \gtrsim 4200$. These corrections are at least 4 orders of magnitude smaller that the linear theory result $C^{(11)}_{\tilde\gamma_{\mr E}}$. 

Note that unlike the results of \citet{CH}, our calculations agree with the expected equivalence between the tangential shear and convergence (cf. Eqs. (\ref{eq:gamma22}, \ref{eq:kappa22}, \ref{eq:gamma13})), as well as between 
cross shear and rotation power spectra (cf. discussion after Eqs. (\ref{eq:gammacross22}, \ref{eq:gamma13})).

\renewcommand{\arraystretch}{2}
\begin{table}[tb]
\caption{$\mathcal O(\Phi^3)$ and $\mathcal O(\Phi^4)$ contributions to the reduced shear E-mode power spectrum.}
\begin{tabular}{l|l|l}
\hline \hline
Type & Term & Contribution to $C_{\tilde{\mr {g}_{\mr E}}}(l)$ \\
\hline\hline

$\mathcal O (\Phi^3)$ reduced shear &$\ensav{\tilde{\gamma}_{\mr E}^{(1)}(\mathbf{l},z_\alpha)\left(\tilde{\gamma}^{(1)}\!\ast\tilde{\kappa}^{(1)}\right)_{\mr E}\!(\mathbf{l''},z_\beta)}$ &
$2\int\!\! \frac{\mr{d}^2l'}{(2\pi)^2}\cos(2\phi_{l'})B_{\kappa}(\mathbf l, \mathbf l',\mathbf{-l-l'};z_\alpha,z_\beta,z_\beta)$ cf. \citep{Shapiro08}\\
\hline

shear with& $\ensav{\tilde{\gamma}_{\mr E}^{(2)}(\mathbf{l}, z_\alpha) \tilde{\gamma}_{\mr E}^{(2)}(\mathbf{l''},z_\beta)} $ &
$4l^2\int\!\! \frac{\mr{d}^2l'}{(2\pi)^2} \left(l'\cos(\phi_{l'})\right)^2
 \left(\mathbf{l'}\cdot\left(\mathbf l - \mathbf{l'}\right)\right)^2 
 M\left(l',|\mathbf l-\mathbf l'|;\chi \right)$\\
 \cline{2-3}
 
multiple deflections & $\ensav{\tilde{\gamma}_{\mr E}^{(1)}(\mathbf{l}, z_\alpha) \tilde{\gamma}_{\mr E}^{(3)}(\mathbf{l''},z_\beta)} $ &
 $-4l^4\int\!\!\frac{\mr d^2 l'}{(2\pi)^2}\left(\mathbf{l}\cdot \mathbf{l'}\right)^2 M(l,l';z_\alpha,z_\beta)$\\
\hline

reduced shear with&$ \ensav{\tilde{\gamma}_{\mr E}^{(1)}(\mathbf{l},z_\alpha)\left(\tilde{\gamma}^{(1)}\!\ast\tilde{\kappa}^{(2)}\right)_{\mr E}\!(\mathbf{l''},z_\beta)}$& 
$-2\int\!\! \frac{\mr{d}^2l'}{(2\pi)^2} \left(\mathbf l\cdot \mathbf{l'}\right)^2 l^2l'^2\cos(2\phi_{l'})$\\
multiple deflections &&$\times \Big\{  \int_0^{\chi_\beta} \!\!\mr d \chi \frac{W(\chi,\chi_\alpha)W(\chi,\chi_\beta)}{\chi^2}\!\!\! 
\int_0^{\chi}\!\! \mr d \chi' \frac{W(\chi',\chi_\beta)W(\chi',\chi)}{\chi'^2}
P_{\Phi}\!\left(l/\chi;z(\chi)\!\right)\!P_{\Phi}\!\left(l'/\chi';z(\chi')\!\right)$\\
&&$ + \int_0^{\chi_\beta} \!\!\mr d \chi \frac{W^2(\chi,\chi_\beta)}{\chi^2}\!\!\! 
\int_0^{\chi}\!\! \mr d \chi' \frac{W(\chi',\chi_\alpha)W(\chi',\chi)}{\chi'^2}
P_{\Phi}\!\left(l'/\chi;z(\chi)\!\right)\!P_{\Phi}\!\left(l/\chi';z(\chi')\!\right)\Big\}$\\
\cline{2-3}

&$\ensav{\tilde{\gamma}_{\mr E}^{(1)}(\mathbf{l},z_\alpha)\left(\tilde{\gamma}^{(2)}\!\ast\tilde{\kappa}^{(1)}\right)_{\mr E}\!(\mathbf{l''},z_\beta)}$&
$-2\int\!\! \frac{\mr{d}^2l'}{(2\pi)^2} \mathbf l\cdot \mathbf{l'}l^2l'^2|\mathbf{l+l'}|$\\
&&$\times \Big\{ 
l \cos(\phi_{\mathbf{l+l'}})
\int_0^{\chi_\beta} \!\!\mr d \chi \frac{W(\chi,\chi_\alpha)W(\chi,\chi_\beta)}{\chi^2}\!\!\! 
\int_0^{\chi}\!\! \mr d \chi' \frac{W(\chi',\chi_\beta)W(\chi',\chi)}{\chi'^2}
P_{\Phi}\!\left(l/\chi;z(\chi)\!\right)\!P_{\Phi}\!\left(l'/\chi';z(\chi')\!\right)$\\
&&$ 
+l' \cos(\phi_{\mathbf{l+l'}}+\phi_{l'})
\int_0^{\chi_\beta} \!\!\mr d \chi \frac{W^2(\chi,\chi_\beta)}{\chi^2}\!\!\! 
\int_0^{\chi}\!\! \mr d \chi' \frac{W(\chi',\chi_\alpha)W(\chi',\chi)}{\chi'^2}
P_{\Phi}\!\left(l'/\chi;z(\chi)\!\right)\!P_{\Phi}\!\left(l/\chi';z(\chi')\!\right)\Big\}$\\
\cline{2-3}
&$\ensav{\tilde{\gamma}_{\mr E}^{(2)}(\mathbf{l},z_\alpha)\left(\tilde{\gamma}^{(1)}\!\ast\tilde{\kappa}^{(1)}\right)_{\mr E}\!(\mathbf{l''},z_\beta)}$&
$-2\int\!\! \frac{\mr{d}^2l'}{(2\pi)^2} \left[ \mathbf{l'}\cdot\left(\mathbf l - \mathbf{l'}\right)\right] |\mathbf l - \mathbf l'|^2 l l'^3 \cos(\phi_{l'})\left\{\cos(2\phi_{l'})+\cos(2\phi_{\mathbf{l-l'}})\right\}$\\
&& $\times \int_0^{\chi_\beta} \!\!\mr d \chi \frac{W(\chi,\chi_\alpha)W(\chi,\chi_\beta)}{\chi^2}\!\!\! 
\int_0^{\chi}\!\! \mr d \chi' \frac{W(\chi',\chi_\beta)W(\chi',\chi)}{\chi'^2}
P_{\Phi}\!\left(l'/\chi;z(\chi)\!\right)\!P_{\Phi}\!\left(|\mathbf{l-l'}|/\chi';z(\chi')\!\right)$\\
\hline

reduced shear with &$\ensav{\tilde{\gamma}_{\mr E}^{(1)}(\mathbf{l},z_\alpha)\left(\tilde{\gamma}^{(1)}\!\ast\tilde{\kappa}^{(1)}\!\ast\tilde{\kappa}^{(1)}\right)_{\mr E}\!(\mathbf{l''},z_\beta)}$
 &$C^{(11)}_{\tilde\gamma_{\mr E}}(l; z_\alpha,z_\beta) \sigma^2_{\tilde\gamma_{\mr E}}( z_\beta)$\\
single deflection &&$+   \int\!\!\frac{\mr d^2 l'}{(2\pi)^2}\cos(2\phi_{l'})\int\!\!\frac{\mr d^2 l'''}{(2\pi)^2}T_{\kappa}(\mathbf l,\mathbf{l'},\mathbf l''',\mathbf{-l-l-l'''}; z_\alpha,z_\beta,z_\beta,z_\beta)$\\
\cline{2-3}

&$\ensav{\left(\tilde{\gamma}^{(1)}\!\ast\tilde{\kappa}^{(1)}\right)_{\mr E}\!(\mathbf{l},z_\alpha)\left(\tilde{\gamma}^{(1)}\!\ast\tilde{\kappa}^{(1)}\right)_{\mr E}\!(\mathbf{l''},z_\beta)}$ & 
$\int\!\!\frac{\mr d^2 l'}{(2\pi)^2}\cos(2\phi_{l'})\Big\{ (\cos(2\phi_{l'})+
\cos(2\phi_{\mathbf{l'-l}}))C^{(11)}_{\tilde\gamma_{\mr E}}(l'; z_\alpha,z_\beta)C^{(11)}_{\tilde\gamma_{\mr E}}(|\mathbf{l'-l}|; z_\alpha,z_\beta)$  \\
 &&$+   \int\!\!\frac{\mr d^2 l'''}{(2\pi)^2}\cos(2\phi_{l'''})T_{\kappa}(\mathbf l',\mathbf{l-l'},\mathbf l''',\mathbf{-l-l'''}; z_\alpha,z_\alpha,z_\beta,z_\beta)\Big\}$\\
 \hline \hline

\end{tabular}
\label{tab:cg}
\end{table}

\renewcommand{\arraystretch}{1}

\subsection{Reduced shear corrections}
\label{sec:cg}
The same methodology used for the corrections to the shear power spectra can also be used to compute the reduced shear terms in Eq.~(\ref{eq:gBig}).
Corrections to the reduced shear power spectra which combine second order and first order distortions contribute through two Wick contractions, for example
\begin{align}
\nonumber \ensav{\tilde{\gamma}_{\mr B}^{(2)}(\mathbf{l,z_\alpha})\left(\tilde{\gamma}^{(1)}\!\ast\tilde{\kappa}^{(1)}\right)_{\mr B}\!\!(\mathbf{l'',z_\beta})} =&\;
-2\epsilon_{IJ}T_I(\mathbf l) \int\!\! \frac{\mr{d}^2l'}{(2\pi)^2} \left[ \mathbf{l'}\cdot\left(\mathbf l - \mathbf{l'}\right)\right] l l' G_J(\mathbf l,\mathbf{l'}) 
\epsilon_{HK}T_H(\mathbf l'')\int\!\! \frac{\mr{d}^2l'''}{(2\pi)^2}l'''^2T_K(\mathbf l''-\mathbf l''') |\mathbf l'' - \mathbf l'''|^2\\
\nonumber  &\times 
\int_0^{\chi_{\beta}}\!\!\! \mr d \chi'' W(\chi'', \chi_{\beta}) \int_0^{\chi_{\beta}}\!\!\! \mr d \chi''' W(\chi''', \chi_{\beta}) 
\ensav{\mathfrak{M}\left(\mathbf{l'},\mathbf{l-l'}; z_\alpha \right)\tilde{\phi}\left(\mathbf l'''; \chi''\right)\tilde{\phi}\left(\mathbf l'' - \mathbf l'''; \chi''' \right)}\\
\nonumber = &-\; (2\pi)^2\delta_{\mr D}(\bfmath l+\bfmath l'')
\;2\!\int\!\! \frac{\mr{d}^2l'}{(2\pi)^2} \left[ \mathbf{l'}\cdot\left(\mathbf l - \mathbf{l'}\right)\right] |\mathbf l - \mathbf l'|^2 l l'^3 \sin(\phi_{l'})\left\{\sin(2\phi_{l'})+\sin(2\phi_{\mathbf{l-l'}})\right\}\\
& \times \int_0^{\chi_\beta} \!\!\mr d \chi \frac{W(\chi,\chi_\alpha)W(\chi,\chi_\beta)}{\chi^2}\!\!\! \int_0^{\chi}\!\! \mr d \chi' \frac{W(\chi',\chi_\beta)W(\chi',\chi)}{\chi'^2}P_{\Phi}\!\left(l'/\chi;z(\chi)\!\right)\!P_{\Phi}\!\left(|\mathbf{l-l'}|/\chi';z(\chi')\!\right),
\end{align}
where we have used $\phi_l = 0$ and $\epsilon_{IJ}T_I(\mathbf l') T_J(\mathbf l'')= \sin(2\phi_{l''}-2\phi_{l'})$.
\renewcommand{\arraystretch}{2}

\begin{table}[tb]
\caption{$\mathcal O(\Phi^4)$ contributions to the reduced shear B-mode power spectrum.}
\begin{tabular}{l|l|l}
\hline \hline
Type & Term & Contribution to $C_{\tilde{\mr {g}_{\mr B}}}(l)$ \\
\hline\hline
shear with &$\ensav{\tilde{\gamma}_{\mr B}^{(2)}(\mathbf{l}, z_\alpha) \tilde{\gamma}_{\mr B}^{(2)}(\mathbf{l''},z_\beta)} $ &
$4l^2\int\!\! \frac{\mr{d}^2l'}{(2\pi)^2} \left(l'\sin(\phi_{l'})\right)^2
 \left(\mathbf{l'}\cdot\left(\mathbf l - \mathbf{l'}\right)\right)^2 
 M\left(l',|\mathbf l-\mathbf l'|;\chi \right)$\\
 multiple deflections & &\\
 \hline
 
reduced shear with& $\ensav{\tilde{\gamma}_{\mr B}^{(2)}(\mathbf{l},z_\alpha)\left(\tilde{\gamma}^{(1)}\!\ast\tilde{\kappa}^{(1)}\right)_{\mr B} \!(\mathbf{l''},z_\beta)}$&
$-2\int\!\! \frac{\mr{d}^2l'}{(2\pi)^2} \left[ \mathbf{l'}\cdot\left(\mathbf l - \mathbf{l'}\right)\right] |\mathbf l - \mathbf l'|^2 l l'^3 \sin(\phi_{l'})\left\{\sin(2\phi_{l'})+\sin(2\phi_{\mathbf{l-l'}})\right\}$\\
multiple deflections & &$\times \int_0^{\chi_\beta} \!\!\mr d \chi \frac{W(\chi,\chi_\alpha)W(\chi,\chi_\beta)}{\chi^2}\!\!\! 
\int_0^{\chi}\!\! \mr d \chi' \frac{W(\chi',\chi_\beta)W(\chi',\chi)}{\chi'^2}
P_{\Phi}\!\left(l'/\chi;z(\chi)\!\right)\!P_{\Phi}\!\left(|\mathbf{l-l'}|/\chi';z(\chi')\!\right)$\\
\hline

reduced shear with & $\ensav{\left(\tilde{\gamma}^{(1)}\!\ast\tilde{\kappa}^{(1)}\right)_{\mr B}\!(\mathbf{l},z_\alpha)\left(\tilde{\gamma}^{(1)}\!\ast\tilde{\kappa}^{(1)}\right)_{\mr B}\!(\mathbf{l''},z_\beta)}$ & 
$\int\!\!\frac{\mr d^2 l'}{(2\pi)^2}\sin(2\phi_{l'})\Big\{ (\sin(2\phi_{l'})+
\sin(2\phi_{\mathbf{l'-l}}))C^{(11)}_{\tilde\gamma_{\mr E}}(l'; z_\alpha,z_\beta)C^{(11)}_{\tilde\gamma_{\mr E}}(|\mathbf{l'-l}|; z_\alpha,z_\beta)$  \\
single deflection & &$+   \int\!\!\frac{\mr d^2 l'''}{(2\pi)^2}\sin(2\phi_{l'''})T_{\kappa}(\mathbf l',\mathbf{l-l'},\mathbf l''',\mathbf{-l-l'''}; z_\alpha,z_\alpha,z_\beta,z_\beta)\Big\}$\\
\hline \hline
\end{tabular}
\label{tab:cgB}
\end{table}

\renewcommand{\arraystretch}{1}

Corrections to the reduced shear power spectra which combine only first order distortions contribute through all Wick contractions plus a connected contribution, for 
example

\begin{align}
\nonumber \ensav{\left(\tilde{\gamma}^{(1)}\!\ast\tilde{\kappa}^{(1)}\right)_{\mr B}\!\!(\mathbf{l}, z_\alpha)
\left(\tilde{\gamma}^{(1)}\!\ast\tilde{\kappa}^{(1)}\right)_{\mr B}\!\!(\mathbf{l''},z_\beta)} =&\; 
\epsilon_{IJ}T_I(\mathbf l)\int\!\! \frac{\mr{d}^2l'}{(2\pi)^2}T_J(\mathbf l') 
\epsilon_{HK}T_H(\mathbf l'')\int\!\! \frac{\mr{d}^2l'''}{(2\pi)^2}T_K(\mathbf l''') \\
&\times \ensav{\tilde{\kappa}^{(1)}(\mathbf l', z_\alpha)\tilde{\kappa}^{(1)}(\mathbf {l-l'}, z_\alpha)
\tilde{\kappa}^{(1)}(\mathbf l''', z_\beta)\tilde{\kappa}^{(1)}(\mathbf {l''-l'''}, z_\beta)}\\
\nonumber =&\;(2\pi)^2\delta_{\mr D}(\bfmath l+\bfmath l'')\int\!\!\frac{\mr d^2 l'}{(2\pi)^2}\sin(2\phi_{l'})\Big\{ (\sin(2\phi_{l'})+\sin(2\phi_{\mathbf{l'-l}}))C^{(11)}_{\tilde\gamma_{\mr E}}(l'; z_\alpha,z_\beta)C^{(11)}_{\tilde\gamma_{\mr E}}(\mathbf{l'-l}; z_\alpha,z_\beta)  \\
& +   \int\!\!\frac{\mr d^2 l'''}{(2\pi)^2}\sin(2\phi_{l'''})T_{\kappa}(\mathbf l',\mathbf{l-l'},\mathbf l''',\mathbf{-l-l'''}; z_\alpha,z_\alpha,z_\beta,z_\beta)\Big\}, 
 \end{align}
where we have omitted a term which only contributes to the $l=0$ mode, and where $T_{\kappa}(\mathbf l_1,\mathbf l_2,\mathbf l_3,\mathbf{-l}_{123}; z_\alpha,z_\alpha,z_\beta,z_\beta)$ is the lensing tomography convergence trispectrum 
\citep{CH01} which we model with the halo model of large scale structure \citep[e.g.,][]{Seljak00,CS02} as summarized in Appendix~\ref{ap:T}. Here, the Gaussian 
contribution, which is 
the dominant term on relevant angular scales, is simply a convolution of the standard $\mathcal O (\Phi^2)$ lensing tomography cross spectra with some geometrical projection factors. Note that in the halo model framework the connected contribution to the B-mode spectrum is downweighted by the geometric projection factors, especially \emph{one-halo} and \emph{(13) two-halo} are strongly suppressed. The connected E-mode terms given in Table~\ref{tab:cg} has opposite angular symmetry and the connected part starts to dominate the signal above $l \sim 8000$.

The analytic expressions for all contributions to the fourth order tangential reduced shear cross spectra are summarized in Table~\ref{tab:cg}. Fig.~\ref{fig:Cg} 
illustrates the numerical values of the different corrections. The fourth order reduced shear corrections of the lensing E-mode power spectrum reach the percent level at small angular scales and hence may be relevant for future weak lensing experiments. Reduced shear generates a small amount of B-mode power, which is about 4 magnitudes smaller than the E-mode signal, and is less than the level of B-mode power generated by observational systematics.

\subsection{Relation between ellipticities and reduced shear}
\label{sc:gestimate}

The linear relation between some measure of image ellipticity and reduced shear (\ref{eq:e}) is only valid in the limit of very weak lensing ($\kappa \ll 1$, 
$|\gamma| \ll1$. In general the relation between image ellipticity and reduced shear depends on the ellipticity measure under consideration. As an example we consider two definitions of the complex image ellipticity here:
\be
\boldsymbol \varepsilon = \frac{1-r}{1+r}e^{2i\phi}
\ee
and
\be
\mathbf e = \frac{1-r^2}{1+r^2}e^{2i\phi},
\ee
where $r\leq 1$ is the minor to major axis ratio of the image, and $\phi$ is the position angle of the major axis.
The latter is frequently employed in observational studies \citep{BJ02}, the former is more 
of theoretical interest due to its simple transformation properties.
The full relation between ellipticity and complex reduced shear $\mathbf g = g_1 + i g_2$ is 
given by
\be
\boldsymbol \varepsilon  = \frac{\boldsymbol \varepsilon ^{(\mr s)}+\mathbf g}{1+\mathbf{g}^*\boldsymbol \varepsilon ^{(\mr s)}}\;\;\;{\rm and}\;\;\;  \mathbf e  = \frac{\mathbf e^{(\mr s)}+2\mathbf g + \mathbf 
g^2\mathbf e^{(\mr s)*}}{1+|\mathbf  g|^2+2\mathcal{R} \left(\mathbf g \mathbf e^{(\mr s)*}\right)},
\ee
where $\mathcal R(\mathbf z)$ is the real part of a complex number $\mathbf z$, $\mathbf e^{(\mr s)}$ and $\boldsymbol \varepsilon ^{(\mr s)}$are the intrinsic ellipticities of the source and where we only consider $|\gamma| <1$, which is certainly true 
for cosmic shear. The linear relation $\ensav{\boldsymbol \varepsilon } =\mathbf  g$ is exact \citep{SS97}, as can be shown using the residue theorem. In the second case, using a Taylor expansion 
\citep{SS95, MandelbaumGG}, the ellipticities can be written as
\be
\ensav{\mathbf e}  = c_1 {\mathbf g} + c_3 |{\mathbf g}|^2{\mathbf g} + \mathcal{O}(g^5) \approx \left(2-\mr e^{(\mr s)^2}\right)\, {\mathbf g}+ \left(-2+5\mr e^{(\mr 
s)^2}-3\mr e^{(\mr s)^4}\right)\, |{\mathbf g}|^2{\mathbf g},
\ee
where $\mr e^{(\mr s)}$ is the absolute value of the intrinsic ellipticity of the source galaxies.  In the practical case of a distribution of intrinsic source ellipticities, one
should replace the powers of $\mr e^{(\mr s)}$ by their moments $\langle \mr e^{(\mr s)\,n}\rangle$.
Shear is typically estimated by taking the mean observed ellipticity $\ensav{\mathbf e}$ and dividing by the response factor $c_1$.
To $\mathcal{O}(\Phi^4)$, this shear estimator reads
\be
\hat{\mathbf g} = \frac{\ensav{\mathbf e}}{c_1} = \mathbf g + \frac{c_3 }{c_1}|{\mathbf g}|^2{\mathbf g}.
\ee
The last term gives rise to one additional contribution to the power spectrum of $\hat{g}_{\mr E}$:
\begin{align}
\nonumber 2\frac{c_3}{c_1}\ensav{\tilde{\gamma}_{\mr E}^{(1)}(\mathbf{l},z_\alpha)\left(\tilde{\gamma}^{(1)}\!\ast\tilde{\gamma}^{(1)}\!\ast\tilde{\gamma}^{(1)}\right)_\mr{E}\!(\mathbf{l'},z_\beta)}
=&2\frac{c_3}{c_1}\; 
\delta_{IJ}T_I(\mathbf{l'})
\int\!\!\frac{\mr d^2 l''}{(2\pi)^2}\int\!\!\frac{\mr d^2 l'''}{(2\pi)^2}
T_H(\mathbf{l'''})T_H(\mathbf{l''})T_J(\mathbf{l'-l''-l'''})\\
\nonumber & \times \Big\langle\tilde{\kappa}^{(1)}(\mathbf l,z_\alpha) \tilde{\kappa}^{(1)}(\mathbf l''',z_\beta)\tilde{\kappa}^{(1)}(\mathbf l'',z_\beta)
\tilde{\kappa}^{(1)}(\mathbf{l'-l''-l'''},z_\beta)\Big\rangle\\
\nonumber = &\;(2\pi)^2\delta_{\mr D}(\bfmath l+\bfmath l')\; 2\frac{c_3}{c_1}\int\!\!\frac{\mr d^2 l''}{(2\pi)^2}\Big\{ (2\cos^2(2\phi_{l''})+1)
C^{(11)}_{\tilde\gamma_{\mr E}}(l; z_\alpha,z_\beta)C^{(11)}_{\tilde\gamma_{\mr E}}(l''; z_\beta,z_\beta)  \\
\nonumber &+  \int\!\!\frac{\mr d^2 l'''}{(2\pi)^2} \cos(2\phi_{l''}-2\phi_{l'''})\cos(2\phi_{\mathbf{-l-l''-l'''}})T_{\kappa}(\mathbf l,\mathbf{l''},\mathbf l''',\mathbf{-l-l''-l'''}; z_\alpha,z_\beta,z_\beta,z_\beta)\Big\}\\
 =(2\pi)^2\delta_{\mr D}(\bfmath l+\bfmath l')\;2\frac{c_3}{c_1}\Big\{
2C^{(11)}_{\tilde\gamma_{\mr E}}(l; z_\alpha,z_\beta) \sigma^2_{\tilde\gamma_{\mr E}}( z_\beta)
&+\int\!\!\frac{\mr d^2 l''\mr d^2 l'''}{(2\pi)^4}\cos(2\phi_{l''}-2\phi_{l'''})\cos(2\phi_{\mathbf{-l-l''-l'''}})T_{\kappa}(\mathbf l,\mathbf{l''},\mathbf l''',\mathbf{-l-l''-l'''}; z_\alpha,z_\beta,z_\beta,z_\beta)\Big\},
\label{eq:NLcorr}
\end{align}
where we have performed the angular integration of the Gaussian contribution in the last step and introduced the shear dispersion
\be
\sigma^2_{\tilde\gamma_{\mr E}}( z_\beta) = \int\!\!\frac{\mr d l'}{2\pi}l' C^{(11)}_{\tilde\gamma_{\mr E}}(l'; z_\beta,z_\beta).
\ee
For the case of the $\boldsymbol \varepsilon$ ellipticity, linearity implies $c_1=1$ and $c_3=0$.  In this case, the correction of Eq.~(\ref{eq:NLcorr}) vanishes.  For the case of 
the ${\mathbf e}$ ellipticity, we have
\be
\frac{c_3}{c_1} = \frac{-2+5\mr e^{(\mr s)^2}-3\mr e^{(\mr s)^4}}{2-\mr e^{(\mr s)^2}}.
\ee
The magnitude of this corrections for the $\mathbf{e}$ ellipticity with $\langle \mr e^{({\mr s})2}\rangle^{1/2}=0.6$ is illustrated in Fig.~\ref{fig:Clb}.

\subsection{Lensing bias corrections}
\label{sec:lensbias}
\begin{figure}
\sidecaption
\includegraphics[width = 0.5\textwidth]{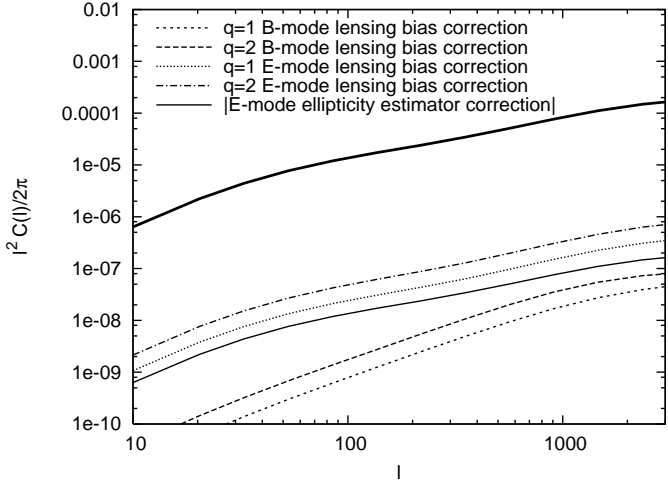}
\caption{Linear order shear power spectrum (thick solid line; Eq. (\ref{eq:firstorder})) and $\mathcal{O}(\Phi^4)$ lensing bias and ellipticity estimator corrections.\newline
The short dashed (dashed) lines show the lensing bias corrections  to the B-mode shear power spectrum (Eq.~\ref{eq:LBB}) assuming q=1 (q=2). The dotted (dashed-dotted) lines show the lensing bias corrections to the E-mode shear power spectrum (Eq.~\ref{eq:LBE}) assuming q=1 (q=2).\newline
The fine solid line illustrates the magnitude of the correction arising from the conversion between ellipticity and reduced shear Eq. (\ref{eq:NLcorr}) for the $\mathbf{e}$ ellipticity with $\langle \mr e^{({\mr s})2}\rangle^{1/2}=0.6$. This correction is negative and its normalization depends on the distribution of source galaxies (see Sect.~\ref{sc:gestimate} for details).\newline
This figure uses the same cosmology and source redshifts as Fig.~\ref{fig:Cg}}
\label{fig:Clb}
\end{figure}
Galaxies are only selected for shear measurement if they are large enough and bright enough to measure their shape. As lensing changes the observed brightness  and size of the lensed galaxies, the number of galaxies selected above some magnitude and size threshold is correlated with the lensing field (this is the well know magnification bias, and the size bias discussed in \citep{FS_size}). Neglecting source clustering, the normalized observed galaxy overdensity due to lensing magnification is given by \citep{FS_3PCF, HGL07}
\be
1+\delta_{\mr{lens}}(\mathbf n) = \frac{1+q\kappa(\mathbf n) + C_1 \kappa^2(\mathbf n)+C_2|\gamma|^2(\mathbf n)}{1+C_1\ensav{\kappa^{2}} +C_2\ensav{|\gamma|^2}}
\approx  1+ q\kappa(\mathbf n) + C_1\left(\kappa^2(\mathbf n) - \ensav{\kappa^2}\right) + C_2\left(|\gamma|^2(\mathbf n) - \ensav{|\gamma|^2}\right)\;,
\ee
where we expanded the magnification to second order\footnote{We note that $\ensav{\kappa} = 0$.  This is because by rotational symmetry the mean deflection angle 
$\ensav{d}=0$, and therefore its derivative $\ensav{\psi_{ij}}=0$.}, 
and where $C_1 = q(q+1)/2$ and $C_2 = q/2$. The parameter $q$ is 
determined by the slope of the luminosity and radius distribution of the sample galaxies and typically $q\sim 1-2$  \citep{FS_size}.

Hence the sampling of the shear field measured from galaxy pairs is modulated by the lensing magnification implying that the observed shear depends on the true shear and the galaxy overdensity
\be
g_I^{\mr{obs}}(\mathbf n) = g_I(\mathbf n)\left(1+\delta_{\mr{lens}}(\mathbf n)\right) 
\approx \gamma_i(\mathbf n)\left\{1+ \kappa(\mathbf n) + \kappa^2(\mathbf n)+ q\kappa(\mathbf n) + q\kappa^2(\mathbf n)
  + C_1[\kappa^2(\mathbf n) - \langle\kappa^2\rangle]
  + C_2[|\gamma(\mathbf n)|^2 - \langle|\gamma|^2\rangle] \right\}\;.
\ee
The standard pair based estimator for the reduced shear correlation functions $\xi_{ab} = \ensav{g_a g_b}$ then becomes \citep[for details see][]{FS_lens}
\be
\ensav{\hat{\xi}_{IJ}(\theta)}= \ensav{\frac{1}{\mathcal N}g_I^{\mr{obs}}(\mathbf n)g_J^{\mr{obs}}(\mathbf n +\mathbf \theta)},
\ee
where $\mathcal N$ is the observed number of galaxy pairs with separation $\theta$ relative to that expected for a random distribution; this is just the  $\frac{DD}{RR}$ correlation function estimator \citep{PH74}.  For large-angle surveys, 
$\mathcal N$ converges to the correlation function,
\be
{\mathcal N} \rightarrow 1 + \langle \delta_{\mr{lens}}(\mathbf n)\delta_{\mr{lens}}(\mathbf n+\mathbf \theta) \rangle.
\ee
Therefore we may write
\be
\ensav{\hat{\xi}_{IJ}(\theta)}=
\frac{\ensav{ g_I^{\mr{obs}}(\mathbf n)g_J^{\mr{obs}}(\mathbf n +\mathbf \theta)}}
{1 + \langle \delta_{\mr{lens}}(\mathbf n)\delta_{\mr{lens}}(\mathbf n+\mathbf \theta) \rangle}.
\label{eq:xiabfull}
\ee
This can be converted to products of correlation functions by conversion to a geometric series,
\be
\ensav{\hat{\xi}_{IJ}(\theta)}= \ensav{ g_I^{\mr{obs}}(\mathbf n)g_J^{\mr{obs}}(\mathbf n +\mathbf \theta)}
\sum_{\upsilon=0}^\infty (-1)^\upsilon \langle \delta_{\mr{lens}}(\mathbf n)\delta_{\mr{lens}}(\mathbf n+\mathbf \theta) \rangle^\upsilon;
\ee
we then note that the $\upsilon$ term in this expansion is of order ${\mathcal O}(\Phi^{2+2\upsilon})$.  Since $\langle\hat{\xi}_{IJ}(\theta)\rangle$ is 
desired to ${\mathcal O}(\Phi^4)$, it suffices to keep only the $\upsilon=0$ and $\upsilon=1$ terms.  Moreover, in the $\upsilon=1$ term, we only require the 
lowest-order expansion of the correlation function $\langle \delta_{\mr{lens}}(\mathbf n)\delta_{\mr{lens}}(\mathbf n+\mathbf \theta) \rangle$, i.e.
\be
\langle \delta_{\mr{lens}}(\mathbf n)\delta_{\mr{lens}}(\mathbf n+\mathbf \theta) \rangle = q^2 \langle \kappa(\mathbf n)\kappa(\mathbf n +\mathbf \theta)
  \rangle + {\mathcal O}(\Phi^3).
\ee
We also need only the lowest-order expansion of $\ensav{g_I^{\mr{obs}}(\mathbf n)g_J^{\mr{obs}}(\mathbf n +\mathbf \theta)}$ in the $\upsilon=1$ term, i.e. we can 
approximate it as $\langle\gamma_I(\mathbf n)\gamma_J(\mathbf n+\mathbf\theta)\rangle$.
Thus we reduce Eq.~(\ref{eq:xiabfull}) to
\be
\ensav{\hat{\xi}_{IJ}(\theta)}
\approx \ensav{g_I^{\mr{obs}}(\mathbf n)g_J^{\mr{obs}}(\mathbf n +\mathbf \theta)}
-q^2 \ensav{\gamma_I(\mathbf n)\gamma_J(\mathbf n +\mathbf \theta)}\ensav{\kappa(\mathbf n)\kappa(\mathbf n +\mathbf \theta)}\;.
\ee
A straightforward generalization to cross-correlations between different redshift slices gives
\be
\ensav{\hat{\xi}_{IJ}(\theta,z_\alpha,z_\beta)}
\approx \ensav{g_I^{\mr{obs}}(\mathbf n,z_\alpha)g_J^{\mr{obs}}(\mathbf n +\mathbf \theta,z_\beta)}
-q^2 \ensav{\gamma_I(\mathbf n,z_\alpha)\gamma_J(\mathbf n +\mathbf \theta,z_\beta)}
  \ensav{\kappa(\mathbf n,z_\alpha)\kappa(\mathbf n +\mathbf \theta,z_\beta)}\;.
\label{eq:xiab}
\ee

We now turn to practical computation.  The terms involving $\ensav{g_I^{\mr{obs}}(\mathbf n,z_\alpha)g_J^{\mr{obs}}(\mathbf n +\mathbf \theta,z_\beta)}$ are all 
identical to terms that we have calculated previously, except with additional factors of $q$, $q^2$, $C_1$, and/or $C_2$, and hence present no new difficulties. The final subtraction term is the product of two expectation values and hence is different from terms that we have previously considered.  This ``product correction'' can 
be evaluated by noting that its contribution to the observed correlation function is the product of the shear and convergence correlation functions.  In Fourier 
space, this means that its contribution to the power spectrum is the convolution of the shear and convergence power spectra:
\be
\Delta C^{\rm prod}_{\tilde\gamma_I\tilde\gamma_J}({\mathbf l}) =
  -q^2 \int \frac{d^2 l'}{(2\pi)^2} C_{\gamma_I\gamma_J}(l') C_{\kappa\kappa}(|{\mathbf l}-{\mathbf l}'|),
\ee
where all power spectra carry the redshift indices $z_\alpha,z_\beta$.  Specializing to the case where ${\mathbf l}$ is along the $x$ coordinate axis, and recalling 
that the $E$-mode shear and convergence power spectra are equal, we can then 
infer a contribution to the observed $E$-mode power spectrum
\be
\Delta C^{\rm prod}_{\tilde\gamma_{\rm E}}({\mathbf l}) =
  -q^2 \int \frac{d^2l'}{(2\pi)^2} \cos^2(2\phi_{{\mathbf l}'})\,
  C^{(11)}_{\tilde\gamma_{\rm E}}(l') C^{(11)}_{\tilde\gamma_{\rm E}}({|\mathbf l}-{\mathbf l}'|);
\label{eq:dcprod}
\ee
the $B$-mode contribution is similar except for the replacement $\cos^2\rightarrow\sin^2$.

Similar to Eqs. (\ref{eq:gg}, \ref{eq:gBig}), we now expand $\ensav{\tilde{g}_{\mr{E/B}}^{\mr{obs}}(\mathbf l)\tilde{g}_{\mr{E/B}}^{\mr{obs}}(\mathbf l'')}$ 
to find the fourth order power spectrum corrections $\Delta C^{\mr{LB}}_{\tilde\gamma_{\mr {E/B}}}$ which arise from lensing bias 
\beq
\nonumber \Delta C^{\mr{LB}}_{\tilde\gamma_{\mr E}}(l;z_\alpha,z_\beta)&=& (2 \pi)^2\delta_{\mr D}(\mathbf l +\mathbf l'')\Biggl\langle\;
2 q\left[
 \ensav{\tilde{\gamma}^{(1)}_{\mr E}\!(\mathbf{l},z_\alpha)\left(\tilde{\gamma}^{(1)}\!\ast\tilde{\kappa}^{(2)}\right)_{\mr E}\!(\mathbf{l''},z_\beta)}+
 \ensav{\tilde{\gamma}^{(1)}_{\mr E}\!(\mathbf{l},z_\alpha)\left(\tilde{\gamma}^{(2)}\!\ast\tilde{\kappa}^{(1)}\right)_{\mr E}\!(\mathbf{l''},z_\beta)}+
 \ensav{\tilde{\gamma}^{(2)}_{\mr E}\!(\mathbf{l},z_\alpha)\left(\tilde{\gamma}^{(1)}\!\ast\tilde{\kappa}^{(1)}\right)_{\mr E}\!(\mathbf{l''},z_\beta)}
 \right]\\
\nonumber& + &(2 q+q^2) \ensav{\left(\tilde{\gamma}^{(1)}\!\ast\tilde{\kappa}^{(1)}\right)_{\mr E}\!(\mathbf{l},z_\alpha)\left(\tilde{\gamma}^{(1)}\!\ast\tilde{\kappa}^{(1)}\right)_{\mr E}\!(\mathbf{l''},z_\beta)}
\label{eq:LBE} + 2 C_2\left[
\ensav{\tilde{\gamma}_{\mr E}^{(1)}(\mathbf{l},z_\alpha)\left(\tilde{\gamma}^{(1)}\!\ast\tilde{\gamma}^{(1)}\!\ast\tilde{\gamma}^{(1)}\right)_{\mr E}\!(\mathbf{l''},z_\beta)}_{\mr c} + C^{(11)}_{\tilde\gamma_{\mr E}}(l; z_\alpha,z_\beta) \sigma^2_{\tilde\gamma_{\mr E}}( z_\beta)
\right]\\
\nonumber &+& 2 q \ensav{\tilde{\gamma}_{\mr E}^{(1)}(\mathbf{l},z_\alpha)\left(\tilde{\gamma}^{(1)}\!\ast\tilde{\kappa}^{(1)}\!\ast\tilde{\kappa}^{(1)}\right)_{\mr E}\!(\mathbf{l''},z_\beta)}+ 2 C_1 \ensav{\tilde{\gamma}_{\mr E}^{(1)}(\mathbf{l},z_\alpha)\left(\tilde{\gamma}^{(1)}\!\ast\tilde{\kappa}^{(1)}\!\ast\tilde{\kappa}^{(1)}\right)_{\mr E}\!(\mathbf{l''},z_\beta)}_{\mr c} \Biggr\rangle_{\alpha \beta}\\
& -&q^2\int\!\!\frac{\mr d^2 l'}{(2\pi)^2} \cos^2(2\phi_{{\mathbf l}'}) C^{(11)}_{\tilde\gamma_{\mr E}}(l'; z_\alpha,z_\beta)C^{(11)}_{\tilde\gamma_{\mr E}}(|\mathbf{l'-l}|; z_\alpha,z_\beta)\,,
\eeq
and
%
%
\beq
\label{eq:LBB} \nonumber \Delta C^{\mr{LB}}_{\tilde\gamma_{\mr B}}(l;z_\alpha,z_\beta) &=& (2 \pi)^2\delta_{\mr D}(\mathbf l +\mathbf l'')\Biggl\langle\;
2 q  \ensav{\tilde{\gamma}^{(2)}_{\mr B}\!(\mathbf{l},z_\alpha)\left(\tilde{\gamma}^{(1)}\!\ast\tilde{\kappa}^{(1)}\right)_{\mr B}\!(\mathbf{l''},z_\beta)}
+(2 q+q^2) \ensav{\left(\tilde{\gamma}^{(1)}\!\ast\tilde{\kappa}^{(1)}\right)_{\mr B}\!(\mathbf{l},z_\alpha)\left(\tilde{\gamma}^{(1)}\!\ast\tilde{\kappa}^{(1)}\right)_{\mr B}\!(\mathbf{l''},z_\beta)}\Biggr\rangle_{\alpha \beta}\\
& -&q^2\int\!\!\frac{\mr d^2 l'}{(2\pi)^2} \sin^2(2\phi_{{\mathbf l}'}) C^{(11)}_{\tilde\gamma_{\mr E}}(l'; z_\alpha,z_\beta)C^{(11)}_{\tilde\gamma_{\mr E}}(|\mathbf{l'-l}|; z_\alpha,z_\beta)\, .
\eeq
In Eq. (\ref{eq:LBE}) we have simplified the terms  which involve the variance of shear or convergence, e.g. the term in Eq. (\ref{eq:xiab}) which is proportional to $C_1$ becomes
\be
C_1\ensav{\tilde{\gamma}_{\mr E}^{(1)}(\mathbf{l},z_\alpha)\left(\tilde{\gamma}^{(1)}\!\ast\tilde{\kappa}^{(1)}\!\ast\tilde{\kappa}^{(1)}\right)_\mr{E}\!(\mathbf{l'},z_\beta)} - C_1\ensav{\tilde{\gamma}_{\mr E}^{(1)}(\mathbf{l},z_\alpha)\tilde{\gamma}_{\mr E}^{(1)}(\mathbf{l'},z_\beta)}\ensav{\kappa^2( z_\beta)} = C_1\ensav{\tilde{\gamma}_{\mr E}^{(1)}(\mathbf{l},z_\alpha)\left(\tilde{\gamma}^{(1)}\!\ast\tilde{\kappa}^{(1)}\!\ast\tilde{\kappa}^{(1)}\right)_\mr{E}\!(\mathbf{l'},z_\beta)}_{\mr c}\,.
\ee
Here the second term is canceled by the disconnected part of the first first term arising from the Wick contraction $C_1 \ensav{\tilde{\gamma}\tilde{\gamma}}\ensav{\tilde{\kappa}\tilde{\kappa}}$, the two other Wick contractions of this term vanish after azimuthal integration. An explicit expression for the connected term is given in Table \ref{tab:cg}.

For the redshift range and cosmology considered in this paper, the second term and third in Eq. (\ref{eq:LBB}) are the dominant contributions. These terms partily cancel and on scales $l \gtrsim 50$ lensing bias effectively 
increases the B-mode power spectrum by approximately a factor $(1+2q)$, which is smaller than the findings of \citet{FS_lens} who only considered the Gaussian contribution to the second 
term in Eq. (\ref{eq:LBB}). The B-mode signal is largest for small angular scaled and high source redshifts. Assuming $q\leq 2$ and a WMAP5 cosmology \citep{WMAP5}, for sources at $z\leq3$  and in the range $l \leq 10000$ the B-mode power spectrum is suppressed by at least a factor 500 (a factor 3000 for $z\leq1$) compared to the shear E-mode power spectrum.

Lensing bias gives rise to a third order correction discussed by \citet{FS_lens}, which is $q$ times the reduced shear correction analyzed by \citet{Shapiro08}. The fourth order E-mode correction generated by lensing bias  Eq. (\ref{eq:LBE}) is more complicated and we will discuss its impact on the E-mode power spectrum in Sect.~\ref{sec:impact}.

The lensing bias E-mode and B-mode corrections are illustrated in Fig.~\ref{fig:Clb} assuming a source redshift $z_\alpha = z_\beta = 1$. Due to uncertainties in modeling the non-linear clustering of matter on small scales we restrict our analysis to $l\leq 3000$, on these scales the lensing bias corrections are below 1\%.

\section{Impact on future surveys}
\label{sec:impact}

\begin{table}[td]
\caption{$Z$ values for the ${\mathcal O}(\Phi^4)$ corrections for different ellipticity estimators with lensing bias.}
\begin{center}
\begin{tabular}{|c|c|c|c|}
\hline \hline
estimator & $q= 0$&$q=1$&$q=2$\\
\hline \hline
 $\boldsymbol \varepsilon$ & 1.14 & 3.19 & 5.31\\
 \hline
 $\mathbf{e}$, $\langle \mr e^{({\mr s})2}\rangle^{1/2}=0.6$& 0.12 & 2.13 & 4.25\\
 \hline\hline
\end{tabular}
\end{center}
\label{tab:Z}
\end{table}

The corrections derived in Sect.~\ref{sec:corrections} generate a small amount of B-mode power, and have a $\lesssim 1\%$ effect on the ellipticity E-mode power spectrum.  These are well below the error bars of current surveys and therefore have no significant effect on published results.  However, future ``Stage IV'' surveys such as LSST, JDEM, and Euclid will be sensitive to sub-percent effects.  We can quantify the importance of the 
higher order lensing corrections by comparing the corrections to the power spectrum $\Delta C(l;z_\alpha,z_\beta)$ to their covariance matrix.  Quantitatively,
\be
Z = \sqrt{\sum_{l\alpha\beta l'\alpha'\beta'}\{{\rm Cov}^{-1}[C(l;z_\alpha,z_\beta),C(l';z_{\alpha'},z_{\beta'})]\} \Delta C(l;z_\alpha,z_\beta) \Delta C(l';z_{\alpha'},z_{\beta'})}
\label{eq:Zcov}
\ee
represents the number of sigmas at which the corrected and uncorrected power spectra could be distinguished by that survey.  Corrections with $Z\ll 1$ are negligible 
in comparison with statistical errors, whereas corrections with $Z\gg 1$ must be known to high accuracy to make full use of the data set.  We have computed 
Eq.~(\ref{eq:Zcov}) assuming a WMAP5 cosmology \citep{WMAP5} for a model survey with a surface density of 30 galaxies/arcmin$^2$, median redshift $z_{\rm med}=1.1$, 
and sky coverage of 10$^4$ deg$^2$, as appropriate for some of the proposed versions of JDEM.  The power spectra were computed in 14 redshift slices and 12 $l$-bins 
with a maximum multipole of $l_{\rm max}=3000$.  The algorithm for computing the covariance matrix is as described in Appendix A.2.d of the JDEM Figure of Merit Science 
Working Group report \citep{FOMSWG}.  Without lensing bias ($q=0$), we find $Z=1.14$ for the linear ellipticity estimator $\boldsymbol \varepsilon$; for the standard estimator 
${\mathbf e}$ and for an rms ellipticity\footnote{The rms ellipticity here includes both the $+$ and $\times$ components, so it is $\sqrt2$ times the rms per axis.} 
$\langle \mr e^{({\mr s})2}\rangle^{1/2}=0.6$, we find $Z=0.12$. Including the lensing bias corrections from Sect.~\ref{sec:lensbias} increases the significance of the 
corrections as detailed in Table~\ref{tab:Z}.  Note that the table includes only the ${\mathcal O}(\Phi^4)$ corrections, and does {\em not} include the ${\mathcal 
O}(\Phi^3)$ corrections that have previously been considered \citep{Shapiro08, FS_3PCF}.
Thus, the perturbative corrections to the weak lensing approximation are expected to be at the level of $\sim 
1-4\sigma$.  These corrections will have to be taken into account for future surveys, but given that they are only $\sim 1-5\sigma$ and should be accurately 
calculable (either directly via ray-tracing simulations, or by analytic expression in terms of the moments of the density field, which can be determined from $N$-body 
simulations), they should not represent a fundamental difficulty.

\section{Discussion}
\label{sec:discussion}

We have calculated the reduced shear power spectra perturbatively to fourth order in the gravitational potential, accounting for the differences between shear and  reduced shear, 
relaxing the Born approximation, and including lens-lens coupling in the calculation of shear and convergence. The full set of corrections to the reduced shear power spectra are given in Table 
\ref{tab:cg} (E-mode) and Table \ref{tab:cgB} (B-mode).  The ellipticity power spectrum contains additional contributions, Eq.~(\ref{eq:NLcorr}), which arises from the non-linearity of the shear 
estimator and depends on the specific definition of ellipticity used, and Eq.~(\ref{eq:LBE}) which is caused by lensing bias. Through order $\Phi^4$, this is the full set of corrections to the power spectrum arising from the lensing process itself. All corrections have been derived within the Limber approximation, and the analysis of ``12" type multiple-deflection corrections is left for future work. Other 
corrections associated with the source galaxy population, such as source clustering and intrinsic alignments, are not treated in this paper.
We find that, depending on the properties of the source galaxy population and on the type of shear estimator used, these corrections will be at the $\sim 1-5\sigma$ level, and thus should be included in the analysis of future precision cosmology weak lensing experiments.

That said, we caution that there are other areas in which the theory of weak lensing needs work if it is to meet ambitious future goals.
Current fitting formula of the non-linear dark matter power spectrum have an accuracy of about $10\%$ at arcminute scales \citep{sm03} and the uncertainty exceeds $30\%$ for $l>10000$ 
\citep{HH09}, due to this difficulty in modeling the non-linear gravitational clustering angular scales of $l> 3000$ are likely to be excluded from parameter fits to cosmic shear measurements. Utilizing near-future $N$-body simulations 
it will become possible to determine the non-linear dark matter power spectrum with percent level accuracy \citep[e.g.,][]{Coyote1, Coyote2}. However, this does not 
account for the effect of baryons, which will likely be important at halo scales and depend critically on the details of baryonic processes (cooling, feedback) 
involved. Baryons in dark matter halos which are able to cool modify the structure of the dark matter halo through adiabatic contraction 
\citep{Blumenthal86,Gnedin04}, causing deviations of the inner halo profile from the simple NFW form and changing the halo mass - halo concentration relation 
\citep[e.g.][]{Rudd08, Pedrosa09}. The latter can be constrained though galaxy-galaxy lensing \citep{MandelbaumGG}, or 
could be internally self-calibrated in a weak lensing survey via its preferential effect on the small-scale power spectrum \citep{Zentner08}.
Baryons in the intergalactic medium may make up  about $10$\% of the mass in the universe, and if their distribution on Mpc scales has been strongly affected by non-gravitational processes 
then they could pose a problem for precise calculation of the matter power spectrum \citep[see][for an extreme and probably unrealistic example]{LG06}.
  
Given these uncertainties in modeling the non-linear matter distribution and that all the corrections derived in this paper are integrals over the non-linear matter 
power spectrum, bispectrum and trispectrum, we refrain from calculating $\mathcal O(\Phi^5)$ and higher corrections. We expect that the corrections derived in this 
paper are sufficient to model the perturbative relation between the non-linear matter distribution and the lensing distortion in weak lensing surveys for the forseeable future.

\begin{acknowledgement}
E.K. and C.H. are supported by the US National Science Foundation under AST-0807337 and the US Department of Energy under DE-FG03-02-ER40701.
C.H. is supported by the Alfred P. Sloan Foundation.
We thank Wayne Hu, Fabian Schmidt, Peter Schneider and Chaz Shapiro for useful discussions.
\end{acknowledgement}

\bibliographystyle{aa}
\bibliography{lensing.bib}

\begin{appendix}

\section{Halo Model Trispectrum}
\label{ap:T}

The trispectrum $T(\mathbf k_1, \mathbf k_2, \mathbf k_3, \mathbf k_4)$ of the dark matter density contrast $\tilde \delta(\mathbf k)$ is defined as
\be
\ensav{\tilde \delta(\mathbf k_1)\tilde \delta(\mathbf k_2)\tilde \delta(\mathbf k_3)\tilde \delta(\mathbf k_4)}_{\mr c} = (2\pi)^3\delta_{\mr D}(\mathbf k_1+\mathbf k_2+\mathbf k_3+\mathbf k_4)T(\mathbf k_1, \mathbf k_2, \mathbf k_3, \mathbf k_4)\;.
\ee
We model the dark matter trispectrum using the halo approach \citep{Seljak00, CS02}, which assumes that all matter is bound in virialized structures, which are assumed to be biased tracers of the density field. Then the statistics of the density field can be described by the dark matter distribution within halos on small scales, and is dominated by the clustering properties of halos and their abundance on large scales. In this model, the trispectrum splits into four terms, which describe the 4-point correlation within one halo (the \emph{one-halo} term $T^{\mr{1h}}$), and between 2 to 4 halos (\emph{two-, three-, four-halo} term)
\be
T = T^{\mr{1h}}+\left(T^{\mr{2h}}_{22}+T^{\mr{2h}}_{13}\right)+T^{\mr{3h}}+T^{\mr{4h}}\;.
\ee
The \emph{two-halo} term is split into two parts, representing correlations between two or three points in the first halo
and two or one point in the second halo.

As halos are the building blocks of the density field in the halo approach, we need to choose models for their internal structure, abundance and clustering in order to build a model for the trispectrum.
In the following we summarize the main ingredients of our implementation of the halo model convergence trispectrum following \citep{CH01}.

We assume the halo profiles to follow the NFW profile \citep{NFW}
\be
\rho(r,c) = \frac{\Delta_{\mr{vir}}\bar\rho c^2}{3(\ln(1+c)-c/(1+c))}\frac{1}{r/r_{\mr{vir}}\left(1+cr/r_{\mr{vir}}\right)^2}\;,
\ee
where $\Delta_{\mr{vir}}$ and $\bar\rho$ are the density contrast and mean density of the universe at virilization,  and $c(M,z)$ is the halo concentration, which we model using the \citet{Bullock01} fitting formula.
We model the halo abundance using the \citet{ST99} mass function
\be
\frac{\mr d n}{\mr d M} \mr d M= \frac{\bar\rho}{M}f(\nu)\mr d \nu = \frac{\bar\rho}{M}A\left[1+(a\nu)^{-p}\right]\sqrt{a\nu}\exp\left(-\frac{a\nu}{2}\right)\frac{\mr d \nu}{\nu},
\ee
 where $A$ and $p$ are fit parameters, and $\nu$ is the peak height $\nu = \delta_c/(D(z)\sigma(M))$.  $\sigma(M)$ is the rms fluctuation of the present day matter density smoothed over a scale $R = (3M/4\pi\bar\rho)^{1/3}$, and $D(z)$ is the growth factor.
To describe the biased relation between the dark matter halo distribution and the density field, we assume a scale independent bias and use the fitting formula of \citet{ST99}
\be
b(\nu) = 1+\frac{a\nu-1}{\delta_{\mr c}}+\frac{2p}{\delta_{\mr c}(1+(a\nu)^p)},
\label{eq:bias}
\ee
and neglect higher order bias functions ($b_2$, etc.).
Following the notation of \citet{CH01} we introduce
\be
I^\beta_\mu (k_1,\cdots,k_\mu; z) =\int \mr d M \frac{\mr d n}{\mr d M}\left(\frac{M}{\bar\rho}\right)^\mu b_\beta(M)\tilde\rho(k_1,c(M,z))\cdots\tilde\rho(k_\mu,c(M,z))\;,
\ee
which describes the correlation of $\mu$ points within the same halo, and where $b_0 = 1$ and  $b_1$ is given by (\ref{eq:bias}). 
Then 
\begin{align}
T^{\mr{1h}}(k_1,k_2,k_3,k_4; z) = I^0_4(k_1,k_2,k_3,k_4; z)\\
T^{\mr{2h}}_{31}(\mathbf k_1,\mathbf k_2,\mathbf k_3,\mathbf k_4; z) = P_{\mr{lin}}(k_{1})D(z)I^1_3(k_2,k_3,k_4; z)I^1_1(k_1; z) + 3\;\mr{perm.}\\
T^{\mr{2h}}_{22}(\mathbf k_1,\mathbf k_2,\mathbf k_3,\mathbf k_4; z) = P_{\mr{lin}}(k_{12})D(z)I^1_2(k_1,k_2; z)I^1_2(k_3,k_4; z) + 2\;\mr{perm.}\\
T^{\mr{4h}} (\mathbf k_1,\mathbf k_2,\mathbf k_3,\mathbf k_4; z)=T^{\mr{pt}}(\mathbf k_1,\mathbf k_2,\mathbf k_3,\mathbf k_4; z)I^1_1(k_1; z)\cdots I^1_1(k_4; z)\;,
\end{align}
where $\mathbf k_{ab} \equiv \mathbf k_a+\mathbf k_b$. We neglect the \emph{3-halo} term, as it has negligible effect on our calculation, and simplify the \emph{4-halo} term using just the trispectrum given by perturbation theory $T^{\mr{pt}}$ \citep{Fry84}.

Finally the tomographic convergence trispectrum can be written as
\beq
\nonumber T_{\kappa}(\mathbf l_1,\mathbf l_2,\mathbf l_3,\mathbf{-l}_{123}; z_\alpha,z_\beta,z_\gamma,z_\delta) = l_1^2 l_2^2 l_3^2 l_{123}^2\int\!\!\mr d \chi \frac{W(\chi,\chi_\alpha)W(\chi,\chi_\beta)W(\chi,\chi_\gamma)W(\chi,\chi_\delta)}{\chi^6}T_\Phi(\mathbf l_1/\chi,\mathbf l_2/\chi,\mathbf l_3/\chi,\mathbf{-l}_{123}/\chi; z(\chi))\\
= \left(\frac{3}{2}\Omega_m H_0^2\right)^4\int\!\!\mr d \chi \frac{W(\chi,\chi_\alpha)W(\chi,\chi_\beta)W(\chi,\chi_\gamma)W(\chi,\chi_\delta)}{\chi^6}(1+z(\chi))^{4}T(\mathbf l_1/\chi,\mathbf l_2/\chi,\mathbf l_3/\chi,\mathbf{-l}_{123}/\chi; z(\chi))\;,
\eeq
where we have used the Poisson equation to relate the potential trispectrum to the matter density trispectrum.
\end{appendix}

\end{document}